\documentclass[11pt]{article}
\date{May 3, 2018}
\usepackage[usenames,dvipsnames]{color}
\usepackage{amsmath,amsthm,amsfonts,amssymb,multicol,amscd,amsbsy}
\usepackage{graphicx}
\usepackage{booktabs}
\usepackage{array}
\usepackage{enumerate}
\usepackage{url}
\usepackage[nodayofweek]{datetime}
\usepackage[english]{babel}
\usepackage{mathtools}
\usepackage[toc,page]{appendix}
\usepackage{verbatim} 
\usepackage{amsthm} 
\usepackage{enumitem}
\usepackage{enumerate}
\usepackage{bbm} 
\usepackage[normalem]{ulem} 
\usepackage{bm}
\usepackage{hyperref}
\usepackage{latexsym}
\usepackage[font=footnotesize,labelfont=bf]{caption}
\usepackage[margin=1.35in]{geometry}

\DeclarePairedDelimiter\ceil{\lceil}{\rceil}
\DeclarePairedDelimiter\floor{\lfloor}{\rfloor}
\usepackage{authblk}

\newcommand{\be}{\begin}
\newcommand{\e}{\end}
\newcommand{\beq}{\begin{equation}}
\newcommand{\eeq}{\end{equation}}
\newcommand{\beqs}{\begin{equation*}}
\newcommand{\eeqs}{\end{equation*}}
\newcommand{\bal}{\begin{align}}
\newcommand{\eal}{\end{align}}
\newcommand{\bals}{\begin{align*}}
\newcommand{\eals}{\end{align*}}

\renewcommand{\l}{\left}
\renewcommand{\r}{\right}

\newcommand{\set}[1]{\mathbb{#1}}

\newcommand{\curly}[1]{\mathcal{#1}}

\newcommand{\setof}[2]{\left\{ #1\; : \;#2 \right\}}

\newcommand{\R}{\set{R}}
\newcommand{\C}{\set{C}}
\newcommand{\Z}{\set{Z}}

\newcommand{\eps}{\epsilon}

\newcommand{\lam}{\lambda}
\newcommand{\Lam}{\Lambda}
\newcommand{\gam}{\gamma}

\newcommand{\al}{\alpha}

\newcommand{\ind}{\mathbbm{1}}		

\newcommand{\del}{\partial}

\theoremstyle{definition}

\numberwithin{equation}{section}

\theoremstyle{remark}

\def\dotuline{\bgroup
  \ifdim\ULdepth=\maxdimen  
   \settodepth\ULdepth{(j}\advance\ULdepth.4pt\fi
  \markoverwith{\begingroup
  \advance\ULdepth0.08ex
  \lower\ULdepth\hbox{\kern.15em .\kern.1em}%
  \endgroup}\ULon}

\def\dashuline{\bgroup
  \ifdim\ULdepth=\maxdimen  
   \settodepth\ULdepth{(j}\advance\ULdepth.4pt\fi
  \markoverwith{\kern.15em
  \vtop{\kern\ULdepth \hrule width .3em}%
  \kern.15em}\ULon}
\allowdisplaybreaks

\newcommand\blfootnote[1]{%
  \begingroup
  \renewcommand\thefootnote{}\footnote{#1}%
  \addtocounter{footnote}{-1}%
  \endgroup
}

\begin{document}
\title{Spectral gaps of frustration-free spin systems with boundary}

\author[1]{Marius Lemm}
\affil[1]{\small{Department of Mathematics,  Harvard University}}
\author[2]{Evgeny Mozgunov}
\affil[2]{\small{University of Southern California}}

\maketitle

\begin{abstract}
In quantum many-body systems, the existence of a spectral gap above the ground state has far-reaching consequences. In this paper, we discuss ``finite-size'' criteria for having a spectral gap in frustration-free spin systems and their applications.

We extend a criterion that was originally developed for periodic systems by Knabe and Gosset-Mozgunov to systems with a boundary. Our finite-size criterion says that if the spectral gaps at linear system size $n$ exceed an explicit threshold of order $n^{-3/2}$, then the whole system is gapped. The criterion takes into account both ``bulk gaps'' and ``edge gaps'' of the finite system in a precise way. The $n^{-3/2}$ scaling is robust: it holds in 1D and 2D systems, on arbitrary lattices and with arbitrary finite-range interactions. One application of our results is to give a rigorous foundation to the folklore that 2D frustration-free models cannot host chiral edge modes (whose finite-size spectral gap would scale like $n^{-1}$).
  \blfootnote{mlemm@math.harvard.edu$\quad$mvjenia@gmail.com}
\end{abstract}


%

\section{Introduction}

A central question concerning a quantum many-body system is whether it is \emph{gapped} or \emph{gapless}. (By definition, a system is ``gapped'' if the difference between the two lowest eigenvalues of the Hamiltonian remains uniformly bounded away from zero in the thermodynamic limit. Otherwise, it is gapless.)

The importance of this concept stems from the fact that ground states of gapped systems enjoy many useful properties. For instance, they display \emph{exponential decay of correlations} \cite{HK,NS}. In one dimension, more is known: The ground states satisfy an \emph{area law} for the entanglement entropy \cite{AKLV,Hastings} and can be well approximated in polynomial time \cite{Aradnew,Landau}. Moreover, the closing of the spectral gap (as a parameter in the Hamiltonian is varied) indicates a quantum phase transition, and is therefore related to topological order, e.g., via the Lieb-Schultz-Mattis theorem \cite{HastingsLSM,LSM}. To summarize, the existence of a spectral gap is a fortunate occurrence that has far-reaching consequences for low-temperature physics. (It is also rare in a probabilistic sense \cite{M}.) 

It is well-known that the existence of a spectral gap may depend on the imposed boundary conditions, and this fact is at the core of our work. 
\\

In general, the question whether a quantum spin system is gapped or gapless is difficult: In 1D, the Haldane conjecture \cite{H1,H2} (``antiferromagnetic, integer-spin Heisenberg chains are gapped'') remains open after 30 years of investigation. In 2D, the ``gapped versus gapless'' dichotomy is in fact undecidable in general \cite{CPW}, even among the class of translation-invariant, nearest-neighbor Hamiltonians.

This paper studies the spectral gaps of a comparatively simple class of models: 1D and 2D \emph{frustration-free} (FF) quantum spin systems with a non-trivial boundary (i.e., open boundary conditions). (A famous FF spin system is the AKLT chain \cite{AKLT}. One general reason why FF systems arise is that any quantum state which is only locally correlated can be realized as the ground state of an appropriate FF ``parent Hamiltonian''  \cite{add0,add1,add2}.)

Specifically, we are interested in \emph{finite-size criteria} for having a spectral gap in such systems. Let us explain what we mean by this.\\

Let $\gam_m$ denote the spectral gap of the Hamiltonian of interest, when it acts on systems of linear size $m$. Let $\tilde\gam_n$ be the \emph{``local gap''}, i.e., the spectral gap of a subsystem of linear size up to $n$. (We will be more precise later.) The finite-size criterion is a bound of the form
\beq\label{eq:idea}
\gam_m\geq c_n(\tilde\gam_n-t_n),
\eeq
for all $m$ sufficiently large compared to $n$ (say $m\geq 2n$). 

Here $c_n>0$ is an unimportant constant, but the value of $t_n$ (in particular its $n$-dependence) is critical. Indeed, if for some fixed $n_0$, we know from somewhere that $\tilde\gam_{n_0}>t_{n_0}$, then \eqref{eq:idea} gives a \emph{uniform lower bound on the spectral gap} $\gam_m$ for all sufficiently large $m$. Accordingly, we call $t_n$ the ``local gap threshold''.

The general idea to prove a finite-size criterion like \eqref{eq:idea} is that the Hamiltonian on systems of linear size $m$ can be constructed out of smaller Hamiltonians acting on subsystems of linear size up to $n$, and these can be controlled in terms of $\tilde\gam_n$. We emphasize that frustration-freeness is essential for this approach.\\

The first finite-size criterion of the kind \eqref{eq:idea} was proved by Knabe \cite{K} for (nearest-neighbor and FF) spin systems with \emph{periodic boundary conditions}. The argument is inspired by the proof of the spectral gap in the AKLT chain \cite{AKLT}. The criterion applies to 1D and can be extended to special 2D systems. It yields a local gap threshold $t_n$ which satisfies $t_n=\Theta(n^{-1})$ as $n\to\infty$. (Here and in the following, we use $f=\Theta(g)$ to mean $f/g\to \mathrm{const}$.)

Knabe's local gap threshold was improved in \cite{GM} to $t_n=\Theta(n^{-2})$ for (nearest-neighbor and FF) systems that are fully translation-invariant (in particular, they have periodic boundary conditions).\\

The \textbf{main message of our paper} is that 1D and 2D FF models with a \emph{non-trivial boundary}, also satisfy a finite-size criterion, and the local gap threshold is $t_n=\Theta(n^{-3/2})$ as $n\to\infty$ (Theorems \ref{thm:main1} and \ref{thm:2D}). 

For 1D nearest-neighbor chains, we compute the constant in front precisely and obtain $t_n=2\sqrt{6}n^{-3/2}$ (Theorem \ref{thm:main1}). 

We find that the $n^{-3/2}$ scaling is rather robust: it holds in 1D and 2D FF systems with general finite-range interactions. We do not know whether it is optimal; see the remarks at the end of the introduction. The $3/2$ scaling exponent has some significance as it corresponds to KPZ scaling behavior.

In particular, our results imply that in \emph{gapless} FF models in 1D and 2D, \emph{the gap cannot close too slowly}. Namely, if we assume $\gam_n=O(n^{-p})$ for some $p>0$, then necessarily $p\geq 3/2$. This corollary improves a recent upper bound $o\l(\frac{\log(n)^{2+\eps}}{n}\r)$ on the spectral gap of gapless FF models \cite{KL}. The fact that $3/2>1$ implies the folklore that 2D FF models cannot produce chiral edge modes, whose finite-size spectral gaps would scale like $n^{-1}$ (Corollary \ref{cor:chiral}).

\subsection{Applications of finite-size criteria}
We will give a more detailed summary of our results below. Before we do so, we explain two ways in which a finite-size criterion like \eqref{eq:idea} can be useful in applications.

\be{enumerate}
\item It implies that a system is \emph{gapped}, if one can prove that $\tilde\gam_{n_0}>t_{n_0}$ holds for some fixed $n_0$. The latter can be proved in some cases by \emph{exactly diagonalizing} the system for very small values of $n$. 

In general, there exist few mathematical tools for deriving a spectral gap. (An important alternative to finite-size criteria is the martingale method developed by Nachtergaele \cite{Nachtergaele}.) 

\item If we \emph{assume} that the system is \emph{gapless}, i.e., $\liminf_{m\to \infty}\gam_m=0$, then \eqref{eq:idea} implies that the finite-size gap $\tilde\gam_n$ can \emph{never exceed} the local gap threshold $t_n$. This ``contrapositive form'' was used in particular in \cite{BG} to classify the gapped and gapless phases of frustration-free (FF), nearest-neighbor spin-1/2 chains. (Another application is given in \cite{KL}.) Here, we use this consequence of our results to rule out chiral edge modes in 2D FF systems (Theorem \ref{thm:chiral}). 
\e{enumerate}

Knabe's original work \cite{K} combined the finite-size criterion with exact diagonalization to obtain explicit lower bounds on the gap of the AKLT model \cite{AKLT} with periodic boundary conditions. As an immediate corollary of our first main result, Theorem \ref{thm:main1}, we can reprove the famous result from \cite{AKLT} that AKLT chain with \emph{open} boundary conditions is gapped (in fact we obtain a numerical lower bound on the gap).\\ 

Other applications of Theorem \ref{thm:main1} pertain to the recently popular (and frustration-free) Fredkin and Motzkin Hamiltonians that can produce maximal violations of the area law \cite{Bravyietal,LM,MovassaghShor,Salbergeretal,UK,Klichetal,ZK}. The boundary is essential in these models; currently they have no periodic analogs. These applications will be explored in a forthcoming work of the first named author with R.~Movassagh

%

\subsection{Summary of main results}

The paper is comprised of two main parts. Part I is on 1D models and part II is on 2D models.

\subsubsection{Part I: Main results in 1D}
In part I, we prove a finite-size criterion for nearest-neighbor FF spin chains with a non-trivial open boundary. On the Hilbert space $(\C^d)^{\otimes m}$, we consider Hamiltonians
\beq\label{eq:Hmdefn}
H_m:=\sum_{i=1}^{m-1}h_{i,i+1}+\Pi_1+\Pi_m,
\eeq
where $h_{i,i+1}$ are local projections acting on two sites and $\Pi_1,\Pi_m$ are local projections acting on a single site. (Possible generalizations are discussed in Section \ref{sect:gen}.) We assume that $H_m$ is frustration-free, i.e., $\ker H_m\neq \{0\}$. 

Let $\gam_m$ be the spectral gap of $H_m$. Our first main result, Theorem \ref{thm:main1}, is the finite-size criterion
$$
\gam_m\geq c_n(\tilde\gam_n-t_n),
$$
with the local gap threshold 
$$
t_n=2\sqrt{6}n^{-3/2},
$$
and the local gap $\tilde \gam_n$ given by
$$
\tilde\gam_n=\min\{\gam^E_{n-1},\gam^B_n\}.
$$
Here, $\gam_{n-1}^E$ is the smallest edge gap achievable by up to $n-1$ sites and $\gam^B_n$ is the bulk gap on $n$ sites. (See Definition \ref{defn:bulkedge} for the precise formulation.) The nomenclature is such that edge gaps take into account one of the boundary projectors $\Pi_1$ or $\Pi_m$, while bulk gaps do not.

Theorem \ref{thm:main1} is the version most suitable for readers interested in combining our results with \emph{exact diagonalization} of small spin chains (see also Remark \ref{rmk:todo}). 

In Part I, we also provide a strong version of the result, Theorem \ref{thm:main2}, in which the minimum edge gap is replaced by a weighted average. 

\subsubsection{Part II: Main results in 2D}\label{ssect:2Dresults}

In Part II, we study two-dimensional FF spin systems with finite-range interactions.\\

First, we establish a rigorous underpinning for the folklore that \emph{2D FF models cannot produce chiral edge modes}. 

More precisely, our \emph{second main result} proves rigorously that the finite-size gap scaling of 2D FF systems is inconsistent with the finite-size scaling $\Theta(n^{-1})$ of a conformal field theory (see Chapter 11 in \cite{CFTbook}) which is expected to appear generally in the description of chiral edge modes (see \cite{Read} for an argument when all edge modes propagate in the same direction). Chiral edge modes are intimately connected to topological order and appear in a variety of modern condensed-matter systems (perhaps most famously in the quantum hall effect). The precise results are in Theorem \ref{thm:chiral} and Corollary \ref{cor:chiral}.

Let us give some background for the folklore. In the monumental work \cite{Kitaev}, Kitaev introduced and studied the exactly-solvable ``honeycomb model''. Among other things, he showed that the model exhibits a phase in which excitations are non-Abelian anyons and the system is gapless due to chiral edge currents. In Appendix D of \cite{Kitaev}, Kitaev considers the special case of a model defined by \emph{commuting} local terms (which puts it in the frustration-free class) and proves the \emph{absence} of chiral edge currents in that special case. Generalizing this observation leads to the folklore that we now rigorously establish here. 

Somewhat surprisingly, the strong version of the 1D result (Theorem \ref{thm:main2}) is sufficient to prove this.\\


Our \emph{last main result} is a very general finite-size criterion, applicable to 2D FF models with boundary (Theorem \ref{thm:2D}). Previous works established finite-size criteria for 2D nearest-neighbor and \emph{periodic} FF systems on the hexagonal lattice \cite{K} and on the square lattice \cite{GM}.

We extend these results to systems with boundary. We have found a robust argument that allows us to treat \emph{arbitrary finite-range} interactions on \emph{arbitrary 2D lattices} and yields the local gap threshold $t_n=\Theta(n^{-3/2})$.\\

The main difference between a 2D finite-size criterion and the 1D finite-size criteria discussed so far, is that in 2D a ``subsystem of linear size $n$'' may take several forms (whereas in 1D it is always a chain of $n$ sites). We call these 2D subsystems ``patches''.

The key idea for our result is to build up the full Hamiltonian from \emph{rhomboidal patches} of spins (they are balls in the $\ell^1$ graph distance; see Figure \ref{fig:rhomboidalpatch} and Definition \ref{defn:rhomb}). The end result (Theorem \ref{thm:2D}) is again a finite-size criterion of the form
$$
\gam_m\geq c_n(\tilde\gam_n-t_n),
$$
with $t_n=\Theta(n^{-3/2})$ and a local gap $\tilde \gam_n$ equal to the minimal gap of rhomboidal patches which are \emph{macroscopic}, in the sense that they act on $\Theta(n)\times \Theta(n)$ sites. (We go through extra work to restrict to macroscopic patches because the spectral gap on very thin patches will not be a genuine 2D quantity.)\\

We elaborate on the choice of \emph{rhomboidal patches}. This patch choice is the only one we know that solves a technical issue in the proof -- a certain counting property about nearest-neighbor interactions has to be satisfied. (This is further explained in Remark \ref{rmk:W}.)

This patch choice is the key innovation of part II and it allows us to prove a very general result. (The technical issue was solved in \cite{GM} for the special case of periodic nearest-neighbor systems on the square lattice by considering two carefully chosen kinds of patches, shown in Fig.\ 1 of \cite{GM}. The issue is absent for nearest-neighbor interactions on the hexagonal lattice considered by Knabe \cite{K}.)

The proof uses a one-step coarse-graining procedure (which may be of independent interest; see Proposition \ref{prop:2Dcg}) to  arrive at an appropriate nearest-neighbor model in which interactions are labeled by plaquettes. The patch choices from \cite{GM} are no longer viable in this setting. Happily, we find that the rhomboidal patches solve the problem.


\subsubsection{Possible extensions}\label{sect:gen}
We first discuss some more or less trivial generalizations of our results and then we mention an open problem.

\be{itemize}
\item The fact that our Hamiltonians are defined in terms of \emph{projections} is a matter of \emph{normalization} only, thanks to frustration-freeness. Indeed, consider the one-dimensional case for simplicity and suppose that the Hamiltonian was defined in terms of general non-negative operators $\tilde h_{j,j+1}$ whose positive spectrum lies in the interval $[a,b]$. Then we set 
$$
h_{j,j+1}:=I-\Pi_{\ker \tilde h_{j,j+1}},
$$
 where $\Pi_X$ projects onto the subspace $X$. From the spectral theorem, we get $a h_{j,j+1}\leq \tilde h_{j,j+1}\leq b h_{j,j+1}$ and this inequality lifts to the Hamilltonian by frustration-freeness.
 
 In this way, our results apply to general frustration-free Hamiltonians as well.

 \item We assume that the Hamiltonian is ``translation-invariant'' in the bulk, meaning that all local terms are defined in terms of fixed projections. This assumption is not strictly necessary for the argument, i.e., the local interactions may depend non-trivially on $j$ (always assuming that the resulting Hamiltonian is FF). However, in this case, the definition of the ``local gap'' becomes rather complicated and it is unclear to us whether a result in this generality would be useful. A less extreme variant of our setup in one dimension, which may be useful in other contexts, is to replace the boundary term $\Pi_m+\Pi_1$  by a general boundary projector $\tilde h_{m,1}$. (In particular, the periodic case can be considered.)

\item Since the arguments are almost entirely algebraic, the results should extend verbatim to frustration-free \emph{lattice fermion systems}, assuming as usual that each interaction term consists of an even number of fermion operators. (This implies that interaction terms of disjoint support commute.) For background on these models, see, e.g., \cite{NSY}.

\item It may be possible to combine our proof of Theorem \ref{thm:2D} with the ideas in \cite{GM} to obtain a $\Theta(n^{-2})$ local gap threshold for fully translation-invariant 2D systems (including periodic boundary conditions) on arbitrary lattices and with arbitrary finite-range interactions. We leave this question to future work.
\e{itemize}

We close the introduction with an \emph{open problem}: We do not know if the local gap threshold scaling $\Theta(n^{-3/2})$ is optimal among the class of FF systems with a boundary. The method of \cite{GM} for deriving $\Theta(n^{-2})$ does not apply because our Hamiltonians do not commute with translations.

We also mention that the $3/2$ scaling exponent for the finite-size gap corresponds to \emph{KPZ scaling behavior}. It can be observed in some open quantum systems, in particular, in exactly solvable Heisenberg $XXZ$ chains with \emph{non-Hermitian} boundary terms \cite{deGier}.

\subsection{Comparison of different notions of spectral gap}
Let us emphasize a subtle, but important matter of convention: In this paper, we are studying the finite-size gap of the many-body Hamiltonian with boundary. In particular, our distinction between ``gapped'' and ``gapless'' phases is \emph{sensitive to the behavior of the system at the boundary}.

As a consequence, our notion of ``gapped'' differs from other commonly considered ones: (a) finite-size spectral gaps of periodic systems as in \cite{GM,K} -- these correspond to finite-size versions of a ``bulk gap''; (b) spectral gaps of infinite-volume ground states in the GNS sense (see, e.g., \cite{BHNY,BN,BNY}). In some sense, our notion of ``gapped'' is the most sensitive one and therefore the strongest: Compared to (a), it is sensitive to small excitation energies due to (partially or completely) edge-bound modes in the finite-system. (This is reflected in the fact that if the open system is gapped, then the periodic system is necessarily gapped as well \cite{K}.) The perspective (b) allows to discount for small excitation energies in the finite-system that amount to a ground-state degeneracy in the infinite-volume limit (and therefore do not influence the infinite-volume spectral gap).

While this sensitivity of our notion of ``gapped'' may appear like a disadvantage at first sight, it is in fact a virtue for some of our purposes: It allows us to directly study the \emph{excitation energies due to edge effects}: Chiral edge modes in 2D systems will influence the finite-size gap we consider, but not any kind of ``bulk gap'' (either in the sense of (a) or (b)).\\

From these considerations, we see that a finite-size criterion for systems with boundary will need to control, in addition to the ``bulk gap'',  the possibility that (partially or completely) edge-bound modes have arbitrarily small excitation energy. In other words, the system should also have a positive ``edge gap''. In our results, this intuition manifests itself as follows: The ``local gap'' $\tilde\gam_n$ in \eqref{eq:idea} is equal to the \emph{minimum} of an appropriate ``bulk gap'' and ``edge gap'' of the finite-size Hamiltonian. If this minimum ever exceeds the local gap threshold $t_n$, then the Hamiltonian is gapped. (See e.g. Theorem \ref{thm:main1} for a precise statement.) 

The phenomenon that \emph{edge-bound states} can become gapless excitations in the thermodynamic limit, in systems with a positive ``bulk gap'' is known to occur. This was rigorously established within the FF class in \cite{BHNY,BN,BNY}. 

\section{Part I: Finite-size criteria for 1D spin chains}
\subsection{The setup}

Let $d\geq 2$. We consider a quantum spin chain on $m\geq 3$ sites which is described by the Hilbert space $(\C^d)^{\otimes m}$. 

We fix a nonzero projection $P:\C^d\otimes \C^d\to \C^d\otimes \C^d$ and define the nearest-neighbor interactions
\beq\label{eq:hjPdefn}
h_{i,i+1}:=
\begin{cases}
P\otimes I_{3,\ldots,m}, &\textnormal{if } i=1,\\
I_{1,\ldots,i-1}\otimes P\otimes I_{i+2,\ldots,m}, &\textnormal{if } i=2,\ldots,m-2,\\
I_{1,\ldots,m-2}\otimes P, &\textnormal{if } i=m-1.
\end{cases}
\eeq
(Here and in the following, $I$ denotes the identity matrix.) Similarly, we fix two projections $P_L:\C^d\to\C^d$ and $P_R:\C^d\to\C^d$ and let 
$$
\Pi_1:=P_L\otimes I_{2,\ldots,m},\qquad \Pi_m:=I_{1,\ldots,m-1}\otimes P_R.
$$ 

Our results will concern the Hamiltonian
\beq\label{eq:HNdefn}
H_m:=\Pi_1+\Pi_m+\sum_{i=1}^{m-1} h_{i,i+1}.
\eeq

Since all the $h_{i,i+1}$ are described by the same matrix $P$, the Hamiltonian is ``translation-invariant in the bulk''.

In the following, we always make 
\be{ass}\label{ass:FF}
For all $m\geq 2$, $H_m$ is frustration-free, i.e., $\inf\mathrm{spec} H_m=0$.
\e{ass} 

This assumption is guaranteed to hold if it holds that $\mathrm{rank}P \leq \max\{d,d^2/4\}$ and $\max\{\mathrm{rank}P_L,\mathrm{rank}P_R\}\leq \max\{1,d/4\}$ \cite{Movassaghetal}. (In the opposite direction, if the local projectors are sampled according to a probability measure that is absolutely continuous with respect to Haar measure and their ranks exceed $d^2/4$, then the model is frustrated with probability $1$ \cite{Ramisstudent}.)

We write $\gam_m$ for the spectral gap of $H_m$, which by Assumption \ref{ass:FF} is equal to the smallest positive eigenvalue of $H_m$. Our main result in 1D is a finite-size criterion, i.e., a lower bound on $\gam_m$ of the form \eqref{eq:idea}. The local gap $\tilde\gam_n$ appearing in the bound is a minimum taken over the following two kinds of gaps: bulk gaps and edge gaps.

\be{defn}\label{defn:bulkedge}
\be{enumerate}[label=(\roman*)]
\item \emph{Bulk gap.} For $2\leq n\leq m$, let $\gam_n^B$ be the gap of the bulk Hamiltonian
$$
H_n^B:=\sum_{i=1}^{n-1} h_{i,i+1}.
$$

\item \emph{Edge gap.}
For $1\leq n\leq m$, let $\gam_n^L$ and $\gam_n^R$ be the gap of the ``left edge'' and ``right edge'' Hamiltonians
$$
H_n^L:=\Pi_1+\sum_{i=1}^{n-1}h_{i,i+1}, \qquad\quad H_n^R:=\Pi_m+\sum_{i=m-n+1}^{m-1} h_{i,i+1}.
$$
The minimal edge gap is defined by
\beq\label{eq:gamEdefn}
\gam_n^E:=\min\l\{1,\min_{2\leq n'\leq n}\min\{\gam_{n'}^L,\gam_{n'}^R\}\r\}.
\eeq
\e{enumerate}
\e{defn}

\be{rmk}
\be{enumerate}[label=(\roman*)]
\item All the involved Hamiltonians are frustration-free by Assumption \ref{ass:FF}, so their spectral gaps are equal to their smallest positive eigenvalues.
\item Let us comment on the role of $\Pi_1$ and $\Pi_m$. The second part of the Hamiltonian $H_m$, namely $\sum_{i=1}^{m-1} h_{i,i+1}$, has open boundary conditions and therefore already represents a model with a true boundary. The additional boundary projections $\Pi_1$ and $\Pi_m$ allow us to include models with special boundary physics, but they can also be chosen identically zero. (Boundary projections are used, e.g., in the Motzkin and Fredkin Hamiltonians \cite{Bravyietal,MovassaghShor,Klichetal}.) As mentioned before, the proof would allow for more general boundary projections $\tilde h_{m,1}$.

If $\Pi_1=\Pi_m=0$, then the boundary also looks like an open chain (i.e., like the bulk) and all the different notions of gap we just introduced agree: $\gam_n=\gam_n^B=\gam_n^L=\gam_n^R$. Moreover, $\Pi_1=\Pi_m=0$ implies $\gam_2=1$, since $h_{1,2}\neq 0$ is a projection.
\e{enumerate}
\e{rmk}

\subsection{Finite-size criterion for periodic spin chains}
For comparison purposes, we briefly recall the finite-size criterion for \emph{periodic} spin chains \cite{GM,K}.

It concerns the  Hamiltonian
$$
H_m^{\mathrm{per}}:=\sum_{i=1}^{m-1} h_{i,i+1}+h_{m,1},
$$
where $h_{m,1}$ is defined analogously as above (i.e., in terms of $P$). In contrast to $H_m$ from \eqref{eq:HNdefn}, $H_m^{\mathrm{per}}$ has no true boundary because of its \emph{periodic boundary conditions}.\\

The following finite-size criterion was proved in \cite{GM}, improving an earlier result from \cite{K}. For this theorem only, we assume that $H_m^{\mathrm{per}}$ is frustration-free for all $m\geq 2$. We write $\gam_m^{\mathrm{per}}$ for its spectral gap.

\be{thm}[\cite{GM}]\label{thm:GM}
Let $m\geq 3$ and $n\leq m/2-1$. Then, we have 
\beq\label{eq:GM}
\gam_m^{\mathrm{per}}\geq \frac{5}{6}\frac{n^2+n}{n-4}\l(\gam_n^B-\frac{6}{n(n+1)}\r).
\eeq
\e{thm}

Comparing \eqref{eq:GM} with the general form \eqref{eq:idea}, we observe that the relevant local gap here is the gap of the bulk Hamiltonian (i.e., $\tilde\gam_n=\gam_n^B$) and the local gap threshold is $t_n=\frac{6}{n(n+1)}=\Theta(n^{-2})$.

Theorem \ref{thm:GM} implies that if $H_m^B=\sum_{i=1}^{m-1} h_{i,i+1}$ is gapped, then $H_m^{\mathrm{per}}$ is gapped as well. The converse is not true in general: The Hamiltonian with open boundary may be gapless, even if $H_m^{\mathrm{per}}$ is gapped. The reason is that there may be gapless edge modes and whether or not this occurs is addressed by our main results below.

\subsection{Main results in 1D}
We come to our first main result, Theorem \ref{thm:main1}. It provides a finite-size criterion for FF and nearest-neighbor 1D chains which have a true boundary (unlike the periodic chains considered above). 

\be{thm}[Main result 1]\label{thm:main1}
Let $m\geq 8$ and $4\leq n\leq m/2$. We have the bound
\beq\label{eq:main}
\gam_m\geq \frac{1}{2^8\sqrt{6n}}\l(\min\{\gam_n^B,\gam_{n-1}^E\}-2\sqrt{6}n^{-3/2}\r).
\eeq
\e{thm}

\be{rmk}\label{rmk:todo}
\be{enumerate}[label=(\roman*)]
\item The proof also yields a bound for $n=3$, albeit with a different threshold:
\beq
\gam_m\geq 2\l(\min\{\gam_3^B,\gam_{2}^E\}-\frac{1}{2}\r).
\eeq

\item If $\Pi_1=\Pi_m=0$, then \eqref{eq:main} simplifies to
\beq\label{eq:2}
\gam_m\geq \frac{1}{2^8\sqrt{6n}}\l(\min_{1\leq n'\leq n}\gam_{n'}-2\sqrt{6}n^{-3/2}\r).
\eeq
In concrete models with $\Pi_1=\Pi_m=0$, it is not uncommon that the function $n\mapsto \gam_n$ is monotonically decreasing for small values of $n$, so that $\min_{1\leq n'\leq n}\gam_{n'}=\gam_n$.

\item The proof of Theorem \ref{thm:main1} actually yields a different threshold $G(n)$, which behaves like $2\sqrt{6}n^{-3/2}$ \emph{asymptotically} as $n\to \infty$, but is appreciably smaller for small values of $n$. (This should be helpful when combining Theorem \ref{thm:main1} with exact diagonalization.) Namely, in \eqref{eq:main} and \eqref{eq:2}, we may replace the threshold $2\sqrt{6}n^{-3/2}$ with the threshold $G(n)$ defined as follows.
$$
\begin{aligned}
G(n):=&\frac{1+a_n^2b_n }{n-1+n^{3/2}a_n},\\
a_n:=&-\frac{n-1}{n^{3/2}}+\sqrt{\l(\frac{n-1}{n^{3/2}}\r)^2+b_n^{-1}},\\
 b_n:=&\frac{6n^3}{(n-1)(n-2)(n-3)}.
\end{aligned}
$$
We note that, for all $n\geq 4$,
$$
G(n)<\min\l\{\frac{1}{n-1},2\sqrt{6}n^{-3/2}\r\}\qquad \mathrm{and}\quad \lim_{n\to\infty}\frac{G(n)}{2\sqrt{6}n^{-3/2}}=1.
$$
For the convenience of readers interested in combining our results with exact diagonalization, we tabulate the first few values of $G(n)$ below (the numbers are obtained by rounding up $G(n)$).

\be{table}[h!!!]
\be{center}
\be{tabular}{|c|c|c|c|c|c|}
 $G(4)$ & $G(5)$ & $G(6)$&	$G(7)$& $G(8)$ & $G(9)$\\
 \hline
 $0.3246$ & $0.2361$ & $0.1833$&	$0.1484$& $0.1238$ & $0.1056$
 \e{tabular}
\e{center}
\e{table}

\item Theorem \ref{thm:main1} generalizes to 1D FF Hamiltonians with \emph{finite interaction-range}. More precisely, one still obtains a local gap threshold scaling $\Theta(n^{-3/2})$, but the multiplicative constants will change. This generalization can be obtained by following a 1D-variant of the coarse graining procedure in our proof of Theorem \ref{thm:2D} on 2D FF models. Since the 1D case is simpler than the 2D case we present, we leave the details to the reader.  
\e{enumerate}
\e{rmk}

\vspace{10pt}

As we mentioned in the introduction, we envision that the condition $\min\{\gam_n^B,\gam_{n-1}^E\}>2\sqrt{6}n^{-3/2}$ can be verified in concrete models by diagonalizing the finite system exactly for small values of $n$. Recall that the edge gap $\gam^E_{n-1}$, defined in \eqref{eq:gamEdefn}, is actually a minimum of gaps of size up to $n-1$. While finding this minimum may look computationally intensive at first sight, note that the complexity of exact diagonalization typically increases exponentially with the system size. Therefore, computing the edge gap $\gam_{n-1}^E$ should be roughly as computationally intensive as computing the bulk gap $\gam_n^B$.

A direct application of Theorem \ref{thm:main1} with $\Pi_1=\Pi_m=0$ (together with the numerical table in \cite{K}) reproves the famous result that the AKLT spin chain with open boundary conditions is gapped \cite{AKLT}. 

Further applications of Theorem \ref{thm:main1}, to the Motzkin and Fredkin Hamiltonians mentioned in the introduction, are currently in preparation, jointly with R.~Movassagh.\\ 

Since $\gam_{n-1}^E$ is defined as a minimum (recall \eqref{eq:gamEdefn}), the bound \eqref{eq:main} is somewhat unstable in the following sense: The expression $\min\{\gam_{n}^B,\gam^E_{n-1}\}$ may be small only because of a single, unusually small gap $\gam_{n'}^L$ or $\gam_{n'}^R$ with $2\leq n'\leq n-1$.

We can actually remedy this instability: Theorem \ref{thm:main2} below is a strong version of Theorem \ref{thm:main1} in which $\gam_{n-1}^E$ is replaced by a \emph{weighted average} of the edge gaps. As a corollary, we obtain a variant of Theorem \ref{thm:main1}, see \eqref{eq:macroversion}, which only involves ``macroscopic'' spectral gaps. 


For all $n\geq 4$ and all $0\leq j\leq n-2$, we introduce the positive coefficients 
$$
c_j=n^{3/2}+\sqrt{6} ((n-2)j-j^2)>0.
$$

\be{thm}[Strong version of Theorem \ref{thm:main1}]\label{thm:main2}
Let $m\geq 8$ and $4\leq n\leq m/2$. 
We have the bound
\beq\label{eq:main2}
\gam_m\geq \frac{1}{2^8\sqrt{6n}}\l(\min\l\{\gam_n,\min_{0\leq j\leq n-2}\frac{\sum_{k=j}^{n-2} c_{k-j} \min\{\gam^L_{k+1},\gam^R_{k+1}\}}{\sum_{k=j}^{n-2} c_{k-j} }\r\}-2\sqrt{6}n^{-3/2}\r).
\eeq
\e{thm}

\be{rmk}
\be{enumerate}[label=(\roman*)]


\item The local gap threshold $2\sqrt{6}n^{-3/2}$ in Theorem \ref{thm:main2} may be improved to $G(n)$ described in Remark \ref{rmk:todo} (iii).

\item A corollary of Theorem \ref{thm:main2} is that the bound \eqref{eq:2} can be rephrased in terms of ``macroscopic'' (comparable to $n$) spectral gaps alone: The expression $\min_{1\leq n'\leq n}\gam_{n'}-2\sqrt{6}n^{-3/2}$ in \eqref{eq:2} can be replaced by, e.g., 
\beq\label{eq:macroversion}
\min_{\floor{n/2}\leq n'\leq n}\gam_{n'}-4\sqrt{6} n^{-3/2}.
\eeq
 This can be seen by following step 2 in the proof of Theorem \ref{thm:chiral}. 
\e{enumerate}
\e{rmk}

\section{Part II: Spectral gaps of 2D spin systems}
Part II contains two results about 2D FF spin systems with a boundary: 

In Theorem \ref{thm:chiral}, we prove that the finite-size gap scaling is inconsistent with the $\Theta(n^{-1})$ finite-size gap scaling. The latter is expected to occur for systems with chiral edge modes. (This can be seen, e.g., via conformal field theory \cite{CFTbook,Read}, but we also give an elementary argument below.). 

Finally, in Theorem \ref{thm:2D}, we give a general 2D finite-size criterion with a local gap threshold $t_n=\Theta(n^{-3/2}$).

For both theorems, the general setup is similar.  The arguments are robust enough to allow for arbitrary bounded, finite-range interactions on arbitrary 2D lattices.  The results are most conveniently phrased (and proved) using specific boundary shapes: a rectangular boundary for Theorem \ref{thm:chiral} and a discrete rhomboidal boundary for Theorem \ref{thm:2D}.

\subsection{The setup}
In a nutshell, we fix a finite subset $\Lam_0$ of a 2D lattice and we define a FF Hamiltonian by translating a fixed ``unit cell of finite-range interactions'' across $\Lam_0$. 

Since this a rather general setup, its precise formulation requires some notation, and we recommend that readers skip this section on a first reading. For the sake of brevity, we restrict to open boundary conditions. Variants, such as periodic boundary conditions on parts of the boundary, require only minor modifications.\\

\dashuline{Hilbert space.}
We let $\Lam_0$ be a \emph{finite} subset of $\Z^2$ (specifying $\Lam_0$ amounts to specifying a boundary shape). At every site $x\in \Lam_0$, we have a local Hilbert space $\C^d$, yielding the total Hilbert space
$$
\curly{H}_{\Lam_0}:=\bigotimes_{x\in \Lam_0} \C^d.
$$

\be{rmk}[Scope]
Since we require $\Lam_0\subset \Z^2$, it may appear that our results only apply to FF models on the lattice $\Z^2$. We clarify that we only use $\Z^2$ to \emph{label the sites} of our lattice. The edges do not matter; the relevant information are the types of interaction terms in the Hamiltonian. As we discuss below, these do not need to respect the lattice structure of $\Z^2$ at all. (For example, nearest neighbor interactions on the triangular lattice can be implemented by introducing interactions between diagonal corners on $\Z^2$.) The formalism also includes the hexagonal lattice, since each copy of $\C^d$ in $\curly{H}_{\Lam_0}$ may be taken to correspond to two (or more) spins within a fixed unit cell.
\e{rmk}

\dashuline{Hamiltonian.}
We define a ``unit cell of interactions'' which we then translate across $\Lam_0$.\\ 

The Hamiltonian is determined as follows. We take a finite family $\curly{S}$ of subsets $S\subset\Lam_0$ containing the point $0$, and for each $S\in \curly{S}$, we fix a projection $P^S:(\C^d)^{\otimes |S|}\to (\C^d)^{\otimes |S|}$, where $|S|$ is the cardinality of $S$. The projections $\{P^S\}_{S\in\curly{S}}$ define the unit cell of interactions.

Given a point $x\in \Lam_0$ such that the set 
$$
x+S:=\setof{y\in\Z^2}{y-x\in S}
$$ lies entirely in $\Lam_0$, we define an operator $P^S_{x+S}:\curly{H}_{\Lam_0}\to \curly{H}_{\Lam_0}$ by 
$$
P^S_{x+S}:= P^S\otimes I_{\Lam_0\setminus (x+S)}.
$$
That is, $P^S_{x+S}$ acts non-trivially only on the subspace $\bigotimes_{y\in x+S}\C^d$ of the whole Hilbert space $\curly{H}_{\Lam_0}=\bigotimes_{y\in\Lam_0}\C^d$.

We can now define the Hamiltonian by
\beq\label{eq:HLam0defn}
H_{\Lam_0}:=\sum_{x\in \Lam_0} \sum_{\substack{S\in\curly{S}:\\ x+S\subset\Lam_0}} P^S_{x+S}.
\eeq
We say that $H_{\Lam_0}$ is ``translation-invariant in the bulk'' because it is constructed by translating the ``unit cell of interactions'' across $\Lam_0$. By construction, $H_{\Lam_0}$ excludes those interaction terms $P^S_{x+S}$ that would involve sites outside of $\Lam_0$, i.e., we are considering open boundary conditions.\\

In this general framework, we make the following assumption.

\be{defn}
Let $e_1=(1,0)$ and $e_2=(0,1)$.
We write $d$ for the standard $\ell^1$ distance on $\Z^2$,i.e.,
$$
d(a_1e_1+a_2e_2,b_1e_1+b_2e_2):=|a_1-b_1|+|a_2-b_2|.
$$
We write $\mathrm{diam}(X)$ for the diameter of a set $X\subset \Z^2$, taken with respect to $d$.
\e{defn}

\be{ass}[Finite interaction range]\label{ass:FR}
 There exists $R>0$ such that $\mathrm{diam}(S)<R$ for all $S\in\curly{S}$.
\e{ass}


We are now ready to state the main results in 2D.

\subsection{Absence of chiral edge modes in 2D FF systems}
We first state the precise result on spectral gaps (Theorem \ref{thm:chiral}) of 2D FF systems. The bound is an ``effectively one-dimensional'' finite-size criterion, in that the subsystems that enter are smaller boxes which have the same ``height'' as the original box. (Accordingly, Theorem \ref{thm:chiral} can be derived from Theorem \ref{thm:main2}.)

Afterwards, we explain why this result provides a rigorous underpinning for the folklore that 2D FF systems cannot host chiral edge modes.

\subsubsection{Effectively 1D finite-size criterion for 2D systems}
We consider systems defined on an open box of dimension $m_1\times m_2$, where $m_1$ and $m_2$ are positive integers. In the framework from above, this means we take
\beq\label{eq:Lam0defn}
\Lam_0=\Lam_{m_1,m_2}:=\setof{a_1e_1+a_2e_2\in \Z^2}{1\leq a_1\leq m_1 \;\;\mathrm{and}\;\; 1\leq a_2\leq m_2}.
\eeq

\be{ass}\label{ass:chiralFF}
For all positive integers $m_1$ and $m_2$, the Hamiltonian $H_{\Lam_{m_1,m_2}}$ is frustration-free.
\e{ass}

We write $\gam_{(m_1,m_2)}$ for the spectral gap of $H_{\Lam_{m_1,m_2}}$. Our second main result is the following bound on $\gam_{(m_1,m_2)}$.

To clarify the dependence of constants, we use the following notation. If a constant $c$ depends on the parameters $p_1,p_2,\ldots$, but not on any other parameters, we write
$$
c=c(p_1,p_2,\ldots).
$$

\be{thm}[Main result 2]\label{thm:chiral}
Let $m_1$, $m_2$ and $4\leq n\leq m_1/R$ be positive integers. There exists positive constants
\beq\label{eq:cdependence}
C_1=C_1(n,m_2,R,\{P^S\}_{S\in\curly{S}}),\qquad 
C_2=C_2(m_2,R,\{P^S\}_{S\in\curly{S}}),
\eeq
such that
\beq\label{eq:thmchiral}
\gam_{(m_1,m_2)}
\geq 
C_1\l(
\min\limits_{\floor{n/2}\leq l\leq n} \gam_{(lR,m_2)}-C_2 n^{-3/2}\r).
\eeq
\e{thm}

\be{rmk}
\be{enumerate}[label=(\roman*)]
\item Let us explain why we call the bound \eqref{eq:thmchiral} ``effectively one-dimensional'' : The left-hand side features systems of size $m_1\times m_2$ and the right-hand side features systems of sizes, roughly, $nR\times m_2$, so only the first dimension of the box varies. We have a finite-size criterion because we can fix $n$ and send $m_1\to\infty$.
\item A similar result holds if we replace the open box by a cylinder of circumference $m_1$ and height $m_2$. More precisely, writing $\gam^{\mathrm{cyl}}_{(m_1,m_2)}$ for the spectral gap of such a cylinder, we have
\beq\label{eq:cyl}
\gam^{\mathrm{cyl}}_{(m_1,m_2)}
\geq 
C_1\l(
\min\limits_{\floor{n/2}\leq l\leq n} \gam_{(lR,m_2)}-C_2 n^{-3/2}\r).
\eeq
\item As we will see in the proof of Corollary \ref{cor:chiral} below, the two key aspects of the bound \eqref{eq:thmchiral} are that (a) the right-hand side only depends on the spectral gaps of ``macroscopic'' subsystem (i.e., ones whose dimensions are comparable to $nR\times m_2$) and (b) $C_2$ is independent of $n$.
\e{enumerate}
\e{rmk}

\subsubsection{The link to chiral edge modes}
We now explain why Corollary \ref{cor:chiral} provides a rigorous underpinning for the folklore that 2D FF models cannot host chiral edge modes. The precise mathematical result we need is

\be{cor}\label{cor:chiral}
Suppose that there exists an integer $M\geq 1$ such that for all $m_1,m_2\geq M$, there exists a constant
$$
C=C(m_2,R,\{P^S\}_{S\in\curly{S}})>1,
$$
such that either
\beq\label{eq:contradictionass}
C^{-1} m_1^{-1} \leq \gam_{(m_1,m_2)}\leq C m_1^{-1}
\eeq
or 
\beq\label{eq:contradictionass'}
C^{-1} m_1^{-1} \leq \gam_{(m_1,m_2)},\qquad \gam_{(m_1,m_2)}^{\mathrm{cyl}}\leq C m_1^{-1}.
\eeq
Then Theorem \ref{thm:chiral} produces a contradiction.
\e{cor}

\be{proof}
Fix $m_2\geq M$. Let $m_1,n\geq 4$ be integers satisfying $M\leq nR\leq m_1$. First, we assume \eqref{eq:contradictionass}. We apply Theorem \ref{thm:chiral} with such $m_1,n$ to get
$$
\gam_{(m_1,m_2)}
\geq 
C_1\l(
\min\limits_{\floor{n/2}\leq l\leq n} \gam_{(lR,m_2)}-C_2 n^{-3/2}\r).
$$
From \eqref{eq:contradictionass}, we then obtain
\beq\label{eq:contra}
C m_1^{-1}\geq 
C_1\l(
C^{-1}R^{-1} n^{-1}-C_2 n^{-3/2}\r).
\eeq
Keeping $m_2\geq M$ fixed, we can find $n_0\geq M/R+1$ sufficiently large so that $
C^{-1}R^{-1} n_0^{-1}-C_2 n_0^{-3/2}>0.$ Finally, we fix $n=n_0$ and send $m_1\to\infty$ in \eqref{eq:contra} to get the desired contradiction $0\geq C^{-1}R^{-1} n_0^{-1}-C_2 n_0^{-3/2}$.
The proof assuming \eqref{eq:contradictionass'} is completely analogous and uses \eqref{eq:cyl} instead of \eqref{eq:thmchiral}.
\e{proof}

To conclude the argument, it suffices to explain why, if we assume that a finite-range lattice Hamiltonian (i.e., some $H_{\Lam_{m_1,m_2}}$) has chiral edge modes, then the assumption of Corollary \ref{cor:chiral} holds. In other words, we have to explain why the finite-size spectral gaps of chiral edge modes can be bounded above and below by $C(m_2) m_1^{-1}$ as soon as $m_1, m_2\geq M$ for some sufficiently large $M$. 

This statement is, in general, difficult to verify rigorously. Nonetheless, the general line of reasoning is rather convincing and is certainly expected to be correct. Moreover, it can be verified in exactly solvable models, for instance in Kitaev's honeycomb model \cite{Kitaev}.\\ 

There are actually two versions of the heuristic argument.

\emph{Version 1} is more hands-on: We assume that the system under consideration hosts chiral edge modes. These are expected to be 1D quasiparticles with dispersion relation $E(k)\propto k$ and this is rigorously known in many cases (perhaps most famously in the fermionic tight-binding model for graphene where the quasiparticles correspond to Dirac cones in the band structure). 

The dispersion relation 
$$
E(k)\propto k,
$$ 
is of course gapless in the continuum, but considering a finite system, say an open box or cylinder of size $m_1\times m_2$, amounts to \emph{discretizing momentum}. The question thus becomes what is the smallest non-zero value of quasiparticle momentum in the finite-size system. (Once one knows this, the finite-size spectral gap is then obtained by evaluating the dispersion relation $E(k)\propto k$ at that smallest momentum.)

For the following discussion, we always assume that $m_1,m_2\geq M$  where $M$ is sufficiently large that the edges of the considered system are well-separated. (In principle, the size of $M$ for a given system is determined by the penetration depth of the edge modes into the bulk. Generally, edge modes are expected to be exponentially localized near the edge.)

First, consider an open $m_1\times m_2$ \emph{box}. In that case, the quasiparticle modes are supported along the entire boundary and therefore their momenta are discretized into steps of order $(m_1+m_2)^{-1}$. This leads to a finite-size gap of order $(m_1+m_2)^{-1}$. Notice that
$$
\frac{1}{1+m_2}\frac{1}{m_1}\leq \frac{1}{m_1+m_2}\leq \frac{1}{m_1}
$$
and so \eqref{eq:contradictionass} is satisfied for $C=C_0(1+m_2)$ with $C_0$ independent of $m_2$. Here we always assume that $m_1,m_2\geq M$, so that we may indeed restrict our attention to the edge behavior. (Note that the exact finite-size gap may incur corrections which are exponentially small in $m_1,m_2$ -- but for the purpose of proving upper and lower bounds of the appropriate order these corrections can be controlled for $m_1,m_2\geq M$.)

Second, we consider the case of a \emph{cylinder} of height $m_2$ and circumference $m_1$. In this case, the quasiparticles propagate along the circumference and their momenta are thus discretized into steps of order $m_1^{-1}$. By the dispersion relation $E(k)\propto k$, this implies that the finite-size gaps satisfy 
$$
C^{-1}m_1^{-1}\leq \gam_{(m_1,m_2)}^{\mathrm{cyl}}\leq C m_1^{-1},
$$
for some $C>1$ independent of $m_2$ --- always assuming that $m_1,m_2\geq M$. Therefore, \eqref{eq:contradictionass'} is also satisfied.\\

\emph{Version 2} goes via conformal field theory. To avoid introducing heavy terminology, we only mention that chiral edge modes are expected to be described by a 1D conformal field theory. (This was verified, e.g., in \cite{Kitaev}; see also \cite{Read} for a general argument that applies if all modes propagate in the same direction.) 

Once the 1D conformal field theory formalism is available, it is well known that finite-size spectral gaps should be of order $\ell^{-1}$ where $\ell$ is the length of the boundary; see, e.g., Chapter 11 in the textbook \cite{CFTbook}. The considerations from version 1 then easily yield the assumptions \eqref{eq:contradictionass} and \eqref{eq:contradictionass'}.

\subsection{A 2D finite-size criterion}

Our third and last main result (Theorem \ref{thm:2D}) is a genuinely 2D finite-size criterion for 2D FF systems with finite-range interactions. The main message is that the local gap threshold is still $\Theta(n^{-3/2})$, where $n$ is the linear size of the subsystem.

\subsubsection{Rhomboidal patches and statement of the result}
\begin{figure}[t]
\begin{center}
\includegraphics[scale=.25]{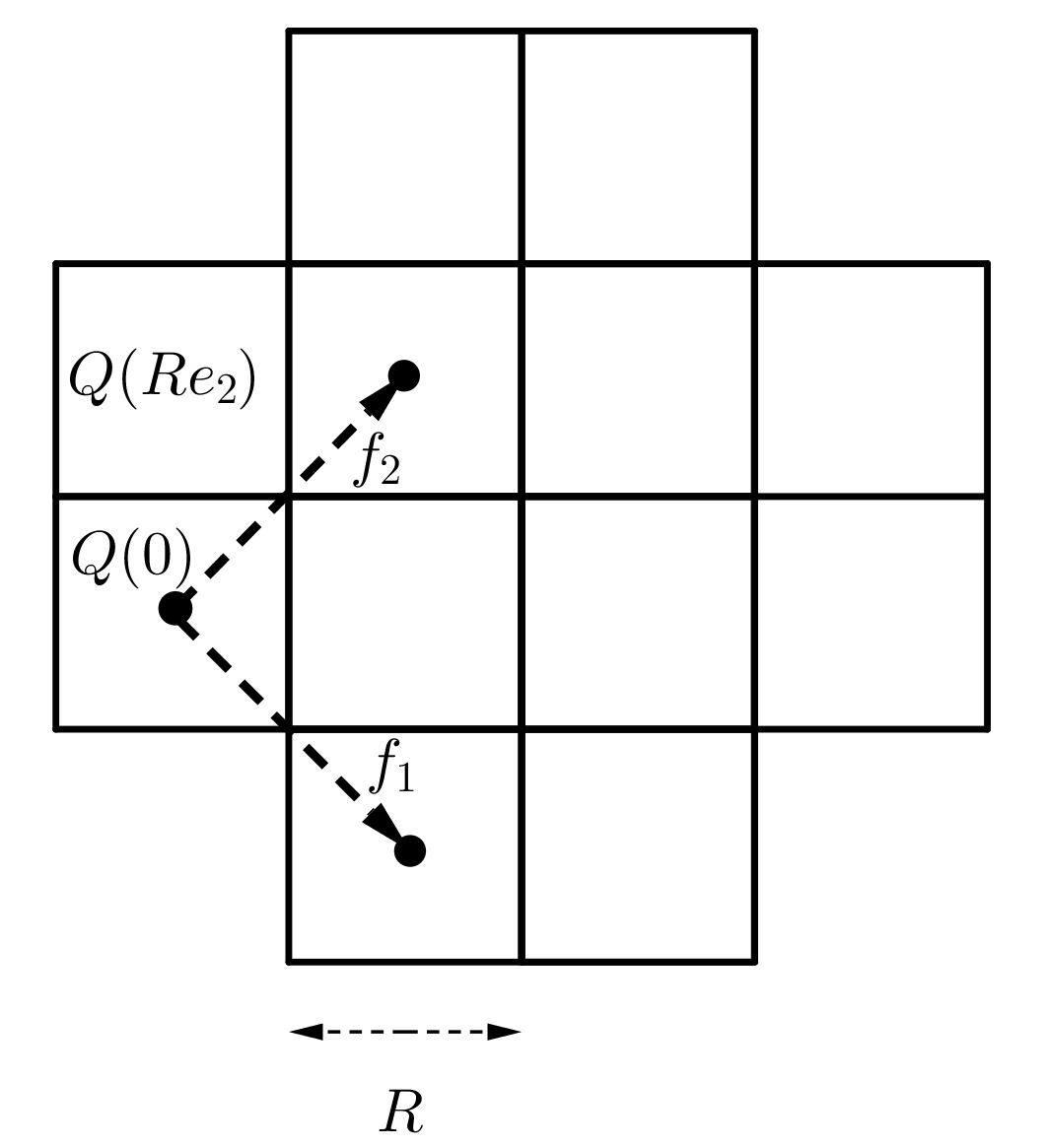}
\end{center}
\caption{The rhomboidal patch $\curly{R}_{2,2}$ from Definition \ref{defn:rhomb}. Each square corresponds to a box $Q(x)$ consisting of $R\times R$ lattice sites ($R$ is also the interaction range). The vectors $f_1$ and $f_2$ are used to generate all the boxes $Q(x)$ in the patch starting from $Q(0)$ and $Q(Re_2)$.}
\label{fig:rhomboidalpatch}
\end{figure}

As we mentioned in the introduction, the key idea is to construct the Hamiltonian out of \emph{rhomboidal patches}. An example is shown in Figure \ref{fig:rhomboidalpatch}. Roughly speaking, rhomboidal patches are rectangles with sides given by the vectors
\beq\label{eq:fdefn}
f_1:=R(e_1-e_2),\qquad f_2:=R(e_1+e_2).
\eeq

Recall that $R>0$ denotes the interaction range (in the $\ell^1$ distance on $\Z^2$), cf.\ Assumption \ref{ass:FR} (i). Without loss of generality, we assume that $R$ is an \emph{odd integer}. 

\be{defn}[Rhomboidal patches]\label{defn:rhomb}
Given $y\in \Z^2$, we call $Q(y)$ the box of sidelength $R$, centered at $y$, i.e.,
$$
Q(y):=\setof{y+a_1e_1+a_2e_2}{|a_1|,|a_2|\leq (R-1)/2}.
$$
Let $n_1$ and $n_2$ be positive integers. We define the rhomboidal patch $\curly{R}_{n_1,n_2}$ by 
\beq\label{eq:defnrhomb}
\curly{R}_{n_1,n_2}:=\bigcup_{j=0}^{n_1-1}\bigcup_{j'=0}^{n_2} Q(j f_1+j'f_2)
\cup \bigcup_{j=0}^{n_1}\bigcup_{j'=0}^{n_2-1} Q(Re_2+j f_1+j'f_2).
\eeq
\e{defn}

The formula \eqref{eq:defnrhomb} is best understood by considering Figure \ref{fig:rhomboidalpatch}.

To each rhomboidal patch $\curly{R}_{n_1,n_2}$ we associate a Hamiltonian $H_{\curly{R}_{n_1,n_2}}$ by using Definition \eqref{eq:HLam0defn} with the choice $\Lam_0=\curly{R}_{n_1,n_2}$, i.e.,
\beq\label{eq:HRdefn}
H_{\curly{R}_{n_1,n_2}}:=\sum_{x\in \curly{R}_{n_1,n_2}} \sum_{\substack{S\in\curly{S}:\\ x+S\subset\curly{R}_{n_1,n_2}}} P^S_{x+S}.
\eeq

\be{ass}\label{ass:convex}
\be{enumerate}[label=(\roman*)]
\item \emph{Interactions are labeled by boxes and lines.} The set of interaction shapes $\curly{S}$ consists only of (a) boxes and (b) lines along $e_1$ or $e_2$. Formally:
$$\begin{aligned}
S\in\curly{S}\quad\Rightarrow\quad &\exists 2\leq r<R\; \textnormal{ such that }\; S=\setof{x\in\Z^2}{d(x,0)\leq r},\\
&\textnormal{or }\; \exists 0<a,b<R\; \textnormal{ s.t. }\; S=\bigcup_{j=-a}^{b} \{je_1\} \textnormal{ or } \bigcup_{j=-a}^{b} \{je_2\}.
\end{aligned}
$$
\item \emph{Frustration-freeness.}
For all positive integers $m_1$ and $m_2$, the Hamiltonian $H_{\curly{R}_{m_1,m_2}}$ is frustration-free.
\e{enumerate}
\e{ass}

We emphasize that assumption (i) is mainly a matter of \emph{convention}: We do \emph{not} assume that the interactions $S$ act non-trivially on the whole box or line -- the interactions should just be \emph{labeled} by such shapes. This labeling can be trivially achieved for any finite-range interaction in the bulk, since we can embed any interaction shape of diameter $<R$ in a box of sidelength $R$. Hence, our theorem below applies to \emph{arbitrary finite-range interactions in the bulk}. 

It is only at the \emph{edge} of $\curly{R}_{n_1,n_2}$ that assumption (i) is marginally restrictive. To explain this further, consider Definition \eqref{eq:HRdefn} of the Hamiltonian $H_{\curly{R}_{n_1,n_2}}$. Notice that the second sum only involves those interaction shapes $S\in\curly{S}$ for which $x+S\subset\curly{R}_{n_1,n_2}$. This condition becomes more restrictive as we increase the interaction shape $S$, and is most restrictive for the box of sidelength $R$ (which labels arbitrary finite-range bulk interactions).

After these preliminaries, we come to our third and last main result. We write $\gam_{\curly{R}_{m_1,m_2}}$ for the spectral gap of $H_{\curly{R}_{m_1,m_2}}$.

\be{thm}[Main result 3]\label{thm:2D}
Let $m_1$, $m_2$ be positive integers. Let $n$ be an even integer such that $2\leq n\leq \min\{m_1/2,m_2/2\}$. There exist constants 
$$
C_1=C_1(n,R,\{P^S\}_{S\in\curly{S}}),\qquad C_2=C_2(R,\{P^S\}_{S\in\curly{S}}),
$$
such that
\beq\label{eq:thm2D}
\gam_{\curly{R}_{m_1,m_2}}\geq C_1\l(\min\limits_{l_1,l_2\in [n/2,n]\cap \Z} \gam_{\curly{R}_{l_1,l_2}} -C_2 n^{-3/2}\r).
\eeq
\e{thm}

\be{rmk}
\be{enumerate}[label=(\roman*)]
\item The constants $C_1$ and $C_2$ are explicit; the more important one $C_2$ is defined in \eqref{eq:see}. They depend on the constants in Proposition \ref{prop:2Dcg} and these can be determined on a case-by-case basis.
\item 
By following the considerations after Theorem \ref{thm:chiral}, Theorem \ref{thm:2D} also yields a result on the absence of chiral edge modes. We do not dwell on this, because the proof given by Theorem \ref{thm:chiral} is simpler, and because the rhomboidal boundary conditions are somewhat artificial.
\e{enumerate}
\e{rmk}
\subsubsection{Discussion on boundary conditions}
 The $n$-sized patches out of which we construct the Hamiltonian need to be rhomboidal (for the technical reason explained in Section \ref{ssect:2Dresults}) and therefore the spectral gaps $\gam_{\curly{R}_{n_1,n_2}}$ appear on the right-hand side of \eqref{eq:thm2D}. The fact that they also appear on the \emph{left}-hand side means that the ``global'' boundary shape is also rhomboidal. This latter choice is not strictly necessary and is made only for convenience. 
 
We first explain why we make this choice. The reason is \emph{geometric}: The relevant subsystems on the right-hand side are the \emph{intersections} of the small rhomboidal patches with the global boundary shape. If the global boundary shape is itself a rhomboid, then its intersections with smaller rhomboids are still rhomboids (see Figure \ref{fig:intersect}). So, in this case, the finite-size criterion \eqref{eq:thm2D} features only one kind of shape, which we find aesthetically pleasing.

Nonetheless, Theorem \ref{thm:2D} generalizes to any kind of ``reasonable'' global boundary condition. The difference for general boundary conditions, say a box, is that the local spectral gaps (the ones on the right-hand side of \eqref{eq:thm2D}) are taken over all the patches that arise by \emph{intersecting the rhomboidal patches with the box}. So there may be lots of different patch shapes whose spectral gaps matter. As above, one may restrict attention to the ``macroscopic'' patches of this kind (i.e., ones with sizes comparable to $n$) and the local gap threshold is still $\Theta(n^{-3/2})$. We trust that the interested reader will find it straightforward to generalize our proof to other boundary shapes.

%
%

%

\section{Proof of the main results in 1D}

\subsection{Preliminaries}
The central tool of the proof are the ``deformed subchain Hamiltonians'' in Definition \ref{defn:Adefn} below. The deformations are defined in terms of an appropriate collection $\{c_0,\ldots,c_{n-2}\}$ of positive real coefficients, on which we assume the following.

\be{ass}\label{ass:c}
The positive real numbers $\{c_0,\ldots,c_{n-2}\}$ satisfy the following two properties.
\be{itemize}
\item \emph{Monotonicity until the midpoint:} $c_j\geq c_{j-1}$ for all $1\leq j\leq \frac{n-2}{2}$.
\item \emph{Symmetry about the midpoint:} $c_{j}=c_{n-2-j}$ for all $0\leq j\leq \frac{n-2}{2}$.
\e{itemize}
\e{ass}

From now on, for every $n\geq 3$, we assume that $\{c_0,\ldots,c_{n-2}\}$ is a fixed collection of positive real numbers satisfying Assumption \ref{ass:c}. We will later minimize the gap threshold over all such collections; see Lemma \ref{lm:coeff}.\\ 

It will be convenient to treat the boundary terms as bond interactions. Therefore we add an artificial $0$th site, i.e., we consider $H_m$ on the enlarged Hilbert space $\C^d\otimes (\C^d)^{\otimes m}$. We associate the boundary projections with the edges incident to the $0$th site as follows,
\beq\label{eq:convenientnota2}
h_{0,1}:=\Pi_1,\qquad h_{m,0}:=\Pi_m.
\eeq
Moreover, we view the resulting chain as a periodic ring on $m+1$ sites, i.e., we identify indices mod $m+1$ via
\beq\label{eq:convenientnota1}
h_{m+1+i,i+1}=h_{i,m+1+i+1}=h_{i,i+1}.
\eeq
In particular, it will be convenient to identify the artificial $0$th site with $m+1$, so that 
\beq\label{eq:hpi}
h_{m,m+1}=\Pi_1,\qquad h_{m+1,m+2}=\Pi_m.
\eeq
 This notation allows us to represent $H_m$ from \eqref{eq:HNdefn} succinctly as
$$
H_m=\sum_{i=1}^{m+1} h_{i,i+1}.
$$

\be{defn}\label{defn:Adefn}
Let $1\leq n\leq m/2$ and $1\leq l\leq m+1$. We introduce the subchain Hamiltonians
$$
A_{n,l}:=\sum_{j=l}^{l+n-2} h_{j,j+1}
$$
and their deformations
$$
\begin{aligned}
B_{n,l}:=\sum_{j=l}^{l+n-2} c_{j-l} h_{j,j+1}.
\end{aligned}
$$
\e{defn}

The definition is such that $n$ denotes the number of sites that the subchain operators act on.

We note that these subchain Hamiltonians fall into \emph{two categories}: Those with $m-n+2\leq l\leq m+1$ ``see the edge'' and contain at least one of the special projections $\Pi_1$ or $\Pi_m$, while those with $1\leq l\leq m-n+1$ do not. Heuristically, the former (latter) describe the behavior of the Hamiltonian $H_m$ at the edge (in the bulk).

\subsection{Proof strategy}
The general strategy to derive a lower bound on the gap of a frustration-free Hamiltonian goes back to Knabe \cite{K}: One proves a lower bound on $H_m^2$ in terms of $H_m$ itself. 

For example, the claimed inequality \eqref{eq:main} is equivalent to the operator inequality
\beq\label{eq:basic}
H_m^2
\geq \frac{1}{2^8\sqrt{6n}}\l(\min\{\gam_n^B,\gam_{n-1}^E\}-2\sqrt{6}n^{-3/2}\r)H_m.
\eeq
We will focus on proving this bound, i.e., Theorem \ref{thm:main1}. The necessary modifications to derive Theorem \ref{thm:main2} are discussed afterwards.\\

The central idea for proving the bound \eqref{eq:basic} is that $H_m^2$ can be related to the sum
\beq\label{eq:Alk}
\sum_{l=1}^{m+1} B_{n,l}^2,
\eeq
and each term in the sum can be bounded from below in terms of the gap of the frustration-free Hamiltonian $B_{n,l}$.

\be{rmk}
Knabe \cite{K} introduced this approach for periodic FF systems, using the undeformed subchain Hamiltonians $A_{n,l}$. Recently, \cite{GM} improved Knabe's bound from $\Theta(n^{-1})$ to $\Theta(n^{-2})$, again for periodic systems. They did so by considering the deformed subchain Hamiltonians $B_{n,l}$, which was motivated by a calculation of Kitaev. Here we adapt the method to systems that have a boundary. The reason why we do not obtain an $\Theta(n^{-2})$ threshold as \cite{GM} did in the periodic case is that their proof requires a strictly translation-invariant Hamiltonian; see Lemma 4 in \cite{GM}.
\e{rmk}



\subsection{Step 1: Relating $H_m^2$ to \eqref{eq:Alk}}
Proposition \ref{prop:rewrite} below gives the desired relation between $H_m^2$ and the sum of deformed subchain Hamiltonians \eqref{eq:Alk}. To prepare for it, we introduce some notation. We write
$$
\{X,Y\}:=XY+YX
$$
for the anticommutator of two operators $X$ and $Y$, and we abbreviate
$$
h_i:=h_{i,i+1}.
$$
We also define the periodic distance function
\beq
d(i,i'):=\min\{|i-i'|,m+1-|i-i'|\}
\eeq
and the matrices
\beq\label{eq:QRdefn}
\qquad Q:=\sum_{i=1}^{m+1} \{h_i,h_{i+1}\},\qquad\quad F:=\sum_{\substack{1\leq i,i'\leq m+1\\ d(i,i')\geq 2}} h_ih_{i'}.
\eeq 

Using this notation, we can relate $H_m^2$ to the sum \eqref{eq:Alk}. On the one hand, a direct computation using $h_i^2=h_i$ (note that this also holds for $i=m$ and $i=m+1$) shows that
\beq\label{eq:Hsquared}
H_m^2=H_m+Q+F.
\eeq
On the other hand, we have

\be{prop}[Knabe-type bound in 1D]\label{prop:rewrite}
Let $3\leq n\leq m/2$. We have the operator inequality
\beq\label{eq:prop}
\begin{aligned}
\sum_{l=1}^{m+1} B_{n,l}^2
\leq& \l(\sum_{j=0}^{n-2} c_j^2\r) H_m+\l(\sum_{j=0}^{n-3} c_j c_{j+1}\r)(Q+F).
\end{aligned}
\eeq
\e{prop}

Proposition \ref{prop:rewrite} is proved in the appendix. Implementing the notation \eqref{eq:convenientnota1} and \eqref{eq:convenientnota2} allows us to essentially repeat the proof given in the translation-invariant case \cite{GM}. The only inequality we use is

\be{lm}\label{lm:sign}
Let $1\leq i,i'\leq m+1$ satisfy $d(i,i')\geq 2$. Then, we have the operator inequality 
$$
h_ih_{i'}\geq 0.
$$
\e{lm}

\be{proof}
The operators $h_i\geq 0$ and $h_{i'}\geq 0$ commute when $d(i,i')\geq 2$.
\e{proof}


\subsection{Step 2: Lower bounds on the deformed subchain Hamiltonians}
The next step is to derive a lower bound on each summand $B_{n,l}^2$.

We take care to separate bulk and edge contributions. Recall Definition \ref{defn:bulkedge} of the edge Hamiltonians $H_n^L$ and $H_n^R$, and recall that the ``edge gap'' $\gam^E_n$ defined in \eqref{eq:gamEdefn} is the minimum of their spectral gaps. 

\be{prop}\label{prop:gap}
Let $1\leq n\leq m/2$. 
\be{enumerate}[label=(\roman*)]
\item 
\textbf{Bulk terms.}
For every $1\leq l\leq m-n+1$, we have the operator inequality
\beq\label{eq:bulkgap}
B_{n,l}^2\geq c_0\gam_n^B B_{n,l},
\eeq

\item \textbf{Edge terms.} For every $m-n+2\leq l\leq m+1$, we have the operator inequality
\beq\label{edgegap}
B_{n,l}^2\geq c_0\gam_{n-1}^E B_{n,l}.
\eeq
\e{enumerate}
\e{prop}

%

\be{proof}[Proof of Proposition \ref{prop:gap}]
Let us denote the spectral gap of $A_{n,l}$ ($B_{n,l}$) by $\gam_{n,l}$ ($\tilde\gam_{n,l}$) respectively. Assumption \ref{ass:FF} implies that all $A_{n,l}$ and $B_{n,l}$ are frustration-free. Moreover, we have
$$
\ker A_{n,l}=\ker B_{n,l}=\bigcap_{i=l}^{l+n-2}\ker h_i.
$$

\dashuline{Proof of (i).} Let $1\leq l\leq m-n+1$. By frustration-freeness, \eqref{eq:bulkgap} is equivalent to the estimate
\beq\label{eq:claimgap}
\tilde\gam_{n,l}\geq c_0\gam_n^B.
\eeq
We now prove this. Since the coefficients $c_j$ are all positive and $c_j\geq c_0$ by Assumption \ref{ass:c}, we have the operator inequality $B_{n,l}\geq c_0 A_{n,l}$. Since $\ker A_{n,l}=\ker B_{n,l}$, this implies $\tilde\gam_{n,l}\geq c_0\gam_{n,l}$. Now, $A_{n,l}$ is  unitarily equivalent (via translation) to $H_n^B$ from Definition \ref{defn:bulkedge} and consequently $\gam_{n,l}=\gam_n^B$. This proves \eqref{eq:claimgap} and thus (i).\\ 

\dashuline{Proof of (ii).} Let $m-n+2\leq l\leq m+1$. We separate $B_{n,l}$ into contributions coming from the left and right ends of the chain, i.e.,
\beq\label{eq:BRLdefn}
\begin{aligned}
B_{n,l}&=B_{n,l}^R+B_{n,l}^L,\\
\textnormal{where}\qquad
B^R_{n,l}&=\sum_{j=l}^{m} c_{j-l} h_j,\qquad B^L_{n,l}=\sum_{j=m+1}^{l+n-2} c_{j-l} h_j.
\end{aligned}
\eeq
We observe that the operators $B^R_{n,l}\geq 0$ and $B^L_{n,l}\geq0 $ commute, since $[h_{m},h_{m+1}]=[\Pi_m,\Pi_1]=0$ by \eqref{eq:hpi}. Hence, we have
$
\{B^R_{n,l},B^L_{n,l}\}\geq0,
$ 
which implies 
$$
B_{n,l}^2=(B_{n,l}^R+B_{n,l}^L)^2\geq (B_{n,l}^R)^2+(B_{n,l}^L)^2.
$$

Recalling \eqref{eq:hpi} and Definition \ref{defn:bulkedge} of the edge Hamiltonian $H_n^R$, we observe that
$$
\begin{aligned}
B_{n,l}^R\geq c_0\sum_{j=l}^{m} h_j=c_0\l(\sum_{j=l}^{m-1} h_j+\Pi_m\r)=c_0 H^R_{m-l+1}.
\end{aligned}
$$
Moreover, we have
$$
\ker B_{n,l}^R=\ker H^R_{m-l+1}=\bigcap_{j=l}^{m}\ker h_j.
$$
Hence, the spectral gap of $B_{n,l}^R$ is at least $c_0 \gam^R_{m-l+1}$, where $\gam^R_{m-l+1}$ is the spectral gap of $H^R_{m-l+1}$. By frustration-freeness, this is equivalent to 
$$
(B_{n,l}^R)^2\geq c_0\gam^R_{m-l+1} B_{n,l}^R.
$$
The same line of reasoning yields $(B_{n,l}^L)^2\geq c_0\gam^L_{l-2+n-m} B_{n,l}^L$ and we have shown that
\beq\label{eq:wehaveshown}
B_{n,l}^2\geq c_0\l(\gam^R_{m-l+1} B_{n,l}^R+\gam^L_{l-2+n-m} B_{n,l}^L\r).
\eeq

The last step of the proof is summarized in the following  

\be{lm}\label{lm:gapestimate}
For all $m-n+2\leq l\leq m+1$, we have 
$$
\gam^R_{m-l+1}\geq \gam^E_{n-1},\qquad\quad \gam^L_{l-2+n-m}\geq \gam^E_{n-1}.
$$
\e{lm}

To see that this lemma implies Proposition \ref{prop:gap}, we apply the given estimates to \eqref{eq:wehaveshown}, using that $B_{n,l}^R,B_{n,l}^L\geq 0$. We find
$$
B_{n,l}^2\geq c_0 \gam^E_{n-1}(B_{n,l}^R+B_{n,l}^L)=c_0 \gam^E_{n-1}B_{n,l},
$$
as claimed.\\

Therefore, it remains to prove Lemma \ref{lm:gapestimate}. We only prove the first inequality,
\beq\label{eq:lmclaim}
\gam^R_{m-l+1}\geq \gam^E_{n-1}.
\eeq
The argument for the second inequality is completely analogous. 

We recall Definition \eqref{eq:gamEdefn} of $\gam^E_{n-1}$, i.e.,
$$
\gam_{n-1}^E=\min\l\{1,\min_{2\leq n'\leq n-1}\min\{\gam_{n'}^L,\gam_{n'}^R\}\r\}.
$$
Notice that $1\leq m-l+1\leq n-1$. If $2\leq m-l+1\leq n-1$, then \eqref{eq:lmclaim} follows directly from the definition of $\gam_{n-1}^E$. Now let $m-l+1=1$, so $B_{n,l}^R=\Pi_m$. If $\Pi_m=0$, then \eqref{eq:lmclaim} holds trivially, otherwise, $B_{n,l}$ has spectral gap equal to $1$ (it is a projection) and \eqref{eq:lmclaim} holds by definition of $\gam_{n-1}^E$. This proves Lemma \ref{lm:gapestimate} and hence also Proposition \ref{prop:gap}.
\e{proof} 

\subsection{Step 3: The choice of the coefficients}
We collect the results of steps 1 and 2. Let us introduce the numbers
$$
\al:=\l(\sum_{j=0}^{n-3}c_jc_{j+1}\r)^{-1},\qquad\quad \beta:=\al \l(\sum_{j=0}^{n-2}c_j^2-\sum_{j=0}^{n-3}c_jc_{j+1}\r),
$$
and the effective gap
$$
\mu_n:=\min\{\gam^E_{n-1},\gam^B_n\}.
$$
From Propositions \ref{prop:rewrite} and \ref{prop:gap}, as well as $B_{n,l}\geq 0$, we obtain
\beq\label{eq:asin}
\begin{aligned}
H_m^2+\beta H_m
\geq \al \sum_{l=1}^{m+1} B_{n,l}^2
\geq \al c_0 \mu_n\sum_{l=1}^{m+1} B_{n,l}
=\al c_0 \mu_n \l(\sum_{j=0}^{n-2} c_j\r) H_m.
\end{aligned}
\eeq
The last equality follows by interchanging the order of summation. By frustration-freeness, this implies the lower bound on the spectral gap
\beq
\label{eq:gapLB}
\gam_m\geq  F(n)(\mu_n-G(n)),
\eeq
where we introduced the quantities 
$$
\begin{aligned}
F(n):=\al c_0 \sum_{j=0}^{n-2}c_j,\qquad G(n):=\frac{2c_0^2+\sum_{j=0}^{n-3} (c_j-c_{j+1})^2}{2c_0 \sum_{j=0}^{n-2}c_j}.
\end{aligned}
$$
To obtain the expression for $G(n)$, we used Assumption \ref{ass:c}. It remains to choose an appropriate set of coefficients $\{c_0,\ldots,c_{n-2}\}$. (Notice that the validity of Assumption \ref{ass:c} and the values of $F(n)$ and $G(n)$ do not change if we multiply each $c_j$ by the same positive number.)

\be{lm}\label{lm:coeff}
Let $n\geq 4$. For $0\leq j\leq n-2$ and $x_n>0$, set
\beq\label{eq:choice}
c_j:=n^{3/2}+x_n((n-2)j-j^2).
\eeq
Then:
\be{enumerate}[label=(\roman*)]
\item The collection $\{c_0,\ldots,c_{n-2}\}$ satisfies Assumption \ref{ass:c}.
\item Let $\phi_n:=\frac{1}{3}(n-1)(n-2)(n-3)$. We have
\beq\label{eq:remains}
\begin{aligned}
F(n)\geq &\frac{1}{24}\frac{n-3}{n^{3/2}}\frac{x_n}{(1+x_n)^2}, \\
G(n)=&\frac{2n^3+x_n^2 \phi_n}{2n^3(n-1)+n^{3/2}x_n\phi_n}.
\end{aligned}
\eeq
\e{enumerate}
\e{lm}

\be{rmk}
 The choice of coefficients \eqref{eq:choice}, together with taking $x_n$ as in \eqref{eq:optimalak} below, is \emph{optimal}, in the sense that it minimizes the local gap threshold $G(n)$ for every fixed $n$. This can be seen by using Lagrange multipliers.
\e{rmk}

\be{proof}[Proof of Lemma \ref{lm:coeff}]
Statement (i) is immediate. For statement (ii), we compute the relevant quantities:
$$
\begin{aligned}
\sum_{j=0}^{n-3} (c_j-c_{j+1})^2=x_n^2\phi_n,\qquad
\sum_{j=0}^{n-2}c_j=n^{3/2}(n-1)+\frac{x_n\phi_n}{2}.
\end{aligned}
$$
This already determines $G(n)$. 

To bound $F(n)$ from below, it remains to control $\al$. We use the coarse estimate 
$$
c_j\leq n^{3/2}+x_n \frac{(n-2)^2}{4}\leq 2n(n-1)(1+ x_n).
$$
We obtain
$$
\sum_{j=0}^{n-3}c_jc_{j+1}\leq 4n^3(n-1)(n-2)(1+ x_n)^2,
$$
and this implies the stated lower bound on $F(n)\geq \al n^{3/2} x_n\phi_n/2$.
\e{proof}


\subsection{Proof of Theorem \ref{thm:main1}}
We recall \eqref{eq:gapLB}, which says
$$
\gam_m\geq  F(n)(\mu_n-G(n)).
$$
The case $n=3$ is trivial: We choose $c_0=c_1=1$ and this yields $F(3)=2$ and $G(3)=1/2$ as stated in Remark \ref{rmk:todo} (i).

For $n\geq 4$, we apply Lemma \ref{lm:coeff}. Notice that we may apply lower bounds for $F(n)$ even if $\mu_n-G(n)<0$, because a priori $\gam_m>0$.\\

\dashuline{Optimal threshold.} We apply Lemma \ref{lm:coeff} with the optimal choice of $x_n>0$, i.e., the choice that minimizes the threshold $G(n)$ and is given by
\beq\label{eq:optimalak}
\begin{aligned}
x_n:=&-b_n\l(\frac{n-1}{n^{3/2}}\r)+\sqrt{b_n^2\l(\frac{n-1}{n^{3/2}}\r)^2+b_n},\\
 b_n:=&\frac{6n^3}{(n-1)(n-2)(n-3)}.
\end{aligned}
\eeq
This yields the threshold $G(n)$ discussed in Remark \ref{rmk:todo}(iii), with the lower bound $F(n)\geq\frac{1}{2^8\sqrt{6}}n^{-1/2}$. To see the latter, we note that $x_n$ in \eqref{eq:optimalak} satisfies $x_n\leq \sqrt{6}$ and apply the estimates discussed below.\\

\dashuline{$2\sqrt{6}n^{-3/2}$ threshold.}
To obtain the slightly weaker threshold $2\sqrt{6}n^{-3/2}$ stated in Theorem \ref{thm:main1}, we set $x_n$ equal to the asymptotic value of \eqref{eq:optimalak}, i.e.,
$$
x_n=\sqrt{6}.
$$
Using elementary estimates, we obtain
$$
\begin{aligned}
F(n)\geq& \frac{1}{24}\frac{n-3}{n^{3/2}}\frac{\sqrt{6}}{(1+\sqrt{6})^2}\geq \frac{1}{16}\frac{1}{\sqrt{6}(1+\sqrt{6})^2}n^{-1/2}\geq \frac{1}{2^8\sqrt{6}}n^{-1/2},\\
G(n)
=&\frac{2n^3+6 \phi_n}{2n^3(n-1)+n^{3/2}\sqrt{6}\phi_n}
\leq 2\sqrt{6} n^{-3/2}.
\end{aligned}
$$
This proves Theorem \ref{thm:main1}. 
\qed

\subsection{Proof of Theorem \ref{thm:main2}}
We define
$$
\tilde\mu_n:=\min\l\{\gam_n,\min_{0\leq j\leq n-2}\frac{\sum_{k=j}^{n-2} c_{k-j} \min\{\gam^L_{k+1},\gam^R_{k+1}\}}{\sum_{k=j}^{n-2} c_{k-j} }\r\}.
$$
We will show that in \eqref{eq:asin}, we may replace $\mu_n$ by $\tilde\mu_n$. Theorem \ref{thm:main2} can then be concluded by following the arguments after \eqref{eq:asin} verbatim, replacing $\mu_n$ by $\tilde\mu_n$ throughout.

Therefore, it suffices to prove 

\be{lm}
We have
\beq\label{eq:asin'}
H_m^2+\beta H_m\geq \al c_0 \tilde\mu_n \sum_{j=0}^{n-2} c_j H_m.
\eeq
\e{lm}

This lemma is proved by using the finer estimate \eqref{eq:wehaveshown}.


\be{proof} 
We recall Definition \eqref{eq:BRLdefn}. For $m-n+2\leq l\leq m+1$, we have \eqref{eq:wehaveshown}, i.e.,
$$
B_{n,l}^2\geq c_0\l(\gam_{m-l+1}^R B_{n,l}^R+\gam_{l-1+n-m}^L B_{n,l}^L\r),
$$

By Propositions \ref{prop:rewrite} and \ref{prop:gap}, we get
$$
\begin{aligned}
H_m^2+\beta H_m
\geq& \al \sum_{l=1}^{m+1} B_{n,l}^2\\
\geq& \al c_0 \l(\gam_n\sum_{l=1}^{m-n+1}B_{n,l}+\sum_{l=m-n+2}^{m+1}\l(\gam_{m-l+1}^R B_{n,l}^R+\gam_{l-2+n-m}^L B_{n,l}^L\r)\r).
\end{aligned}
$$
We consider the right edge more closely and find
$$
\begin{aligned}
\sum_{l=m-n+2}^{m+1}\gam_{m-l+1}^R B_{n,l}^R
=&\sum_{j=l}^{m} h_j\l(\sum_{l=m-n+2}^j c_{j-l} \gam^R_{m-l+1}\r)
=\sum_{j=l}^{m} h_j\l(\sum_{k=m-j}^{n-2} c_{k-m+j} \gam^R_{k+1}\r)\\
\geq& \tilde \mu_n \sum_{j=l}^{m} h_j\l(\sum_{k=m-j}^{n-2} c_{k-m+j}\r)
= \tilde \mu_n \sum_{l=m-n+2}^{m+1}\gam_{m-l+1}^R B_{n,l}^R.
\end{aligned}
$$
A similar result holds for the left edge. (There one also uses the symmetry $c_j=c_{n-2-j}$ of the coefficients.) Since $\gam_n\geq \tilde\mu_n$, this proves the lemma and hence Theorem \ref{thm:main2}.
\e{proof}

\section{Proof of the key result concerning chiral edge modes}
In this section, we prove Theorem \ref{thm:chiral} from Theorem \ref{thm:main2}.

Let $m_1$ and $m_2$ be positive integers and recall that
$$
\Lam_0=\Lam_{m_1,m_2}:=\setof{a_1e_1+a_2e_2\in \Z^2}{1\leq a_1\leq m_1 \;\;\mathrm{and}\;\; 1\leq a_2\leq m_2}.
$$

The proof of Theorem \ref{thm:chiral} is based on the following two steps. 

In step 1, we relate the spectral gap of $H_{\Lam_{m_1,m_2}}$ to the spectral gap of an ``effective'' 1D FF nearest-neighbor Hamiltonian (Proposition \ref{prop:cg}). The effective 1D Hamiltonian is obtained by a (rather brutal) one-step \emph{coarse-graining procedure}, in which we group $\Lam_{m_1,m_2}$ into metaspins, each describing the collection of $\C^d$-spins within a rectangle of width $R$ and height $m_2$. 

In step 2, we apply Theorem \ref{thm:main2} to the effective 1D Hamiltonian. The key observation is that in the weighted average over spectral gaps, we may restrict to ``macroscopic chains'', i.e., those of sizes between $n/2$ and $n$. This yields Theorem \ref{thm:chiral}.

\subsection{Step 1: The effective 1D Hamiltonian}
We recall that $\gam_{(m_1,m_2)}$ is the spectral gap of $H_{\Lam_0}$. The basic idea for the one-step coarse-graining procedure is to split the box $\Lam_{m_1,m_2}$ into $m$ disjoint rectangles $\Lam_j$ of dimensions $R\times m_2$ (assuming that $m_1=mR$). See Figure \ref{fig:1dcg} below. That is, we define the rectangle
$$
\Lam_j:=\setof{a_1e_1+a_2e_2\in \Lam_{m_1,m_2}}{1+(j-1)R\leq a_1\leq jR},
$$
for $1\leq j\leq m$. We decompose the Hilbert space in the same way, i.e.,
\beq\label{eq:Hjdefn}
\curly{H}_{\Lam_{m_1,m_2}}=\bigotimes_{a_1=1}^{m_1}\bigotimes_{a_2=1}^{m_2}\C^d
=\bigotimes_{j=1}^{m} \curly{H}_j,\qquad \textnormal{where}\quad \curly{H}_j:=\bigotimes_{x\in \Lam_j}\C^d.
\eeq
Here $\curly{H}_j$ is the Hilbert space corresponding to a rectangle $\Lam_j$. Note that it is isomorphic to $\C^{d^{m_2R}}$. We call each $\curly{H}_j$ a ``metaspin''. 

\begin{figure}[t]
\begin{center}
\includegraphics[scale=.15]{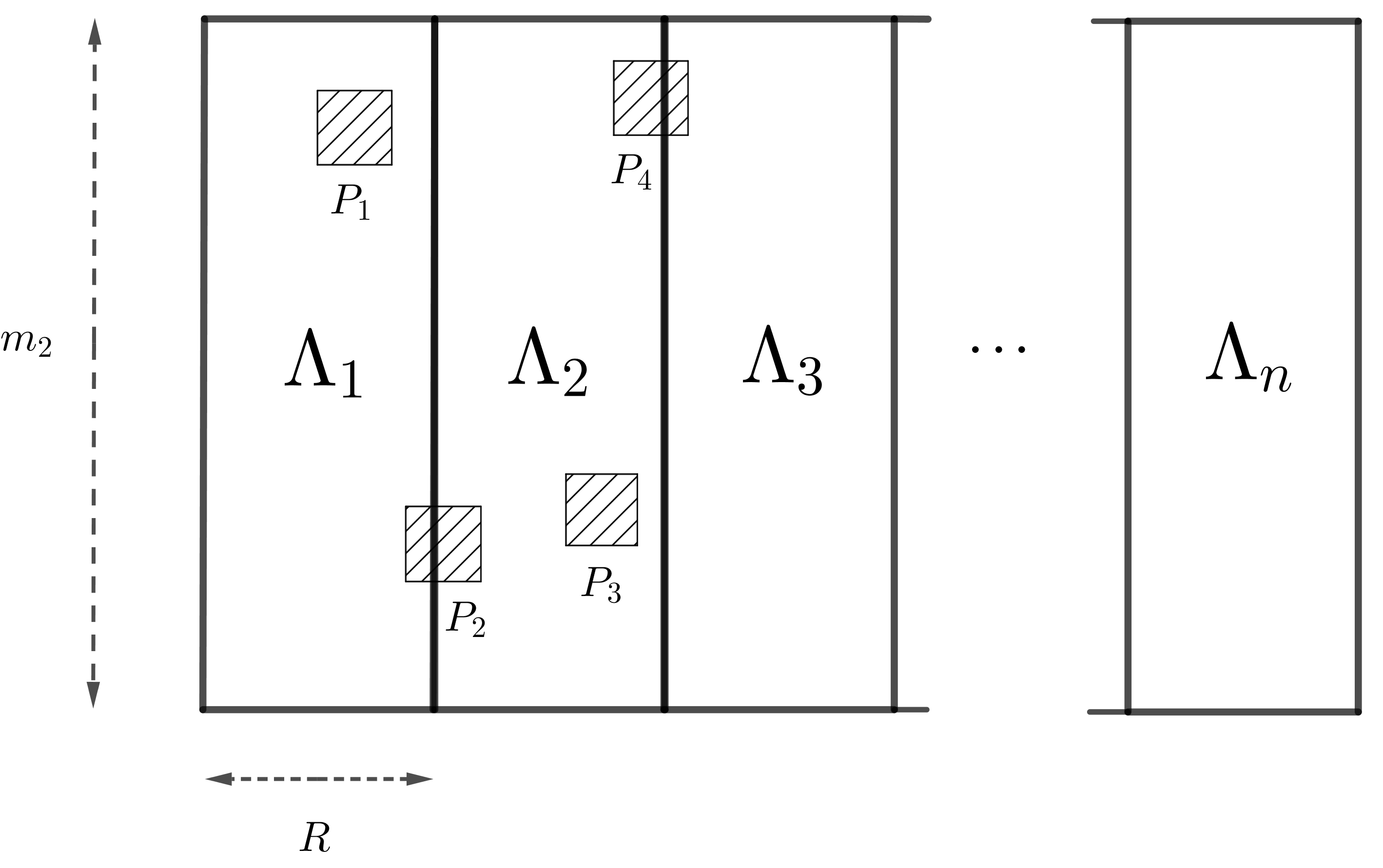}
\end{center}
\caption{The one-step coarse-graining procedure replaces all the spins within an $R\times m_2$ rectangle $\Lam_j$ by a single metaspin. The key fact is that only rectangles that touch can interact. (This makes the effective 1D Hamiltonian nearest-neighbor.) We group the interaction terms into $\tilde h_{j,j+1}$'s (with $H_{\Lam_0}=\sum_{j=1}^{m-1} \tilde h_{j,j+1}$) as follows: Interactions of the type $P_1$ and $P_2$ are in $\tilde h_{1,2}$; interactions of the type $P_3$ are split 50/50 between $\tilde h_{1,2}$ and $\tilde h_{2,3}$; interactions of the type $P_4$ are in $\tilde h_{2,3}$; etc.}
\label{fig:1dcg}
\end{figure}

\be{prop}[Effective 1D Hamiltonian]\label{prop:cg}
Let $m_1$ and $m_2$ be positive integers and let $m_1=mR$ for another integer $m$. There exists a projection $P_{\mathrm{eff}}$ acting on $\C^{d^{m_2R}}\otimes \C^{d^{m_2R}}$ such that the following holds. For each $1\leq j\leq m-1$, we define a projection $h_{j,j+1}$ on $\curly{H}_{\Lam_{m_1,m_2}}$ via \eqref{eq:hjPdefn}, replacing $P$ with $P_{\mathrm{eff}}$. We define the Hamiltonian
$$
H_{m}^{1D}:=\sum_{j=1}^{m-1} h_{j,j+1}
$$
on $\curly{H}_{\Lam_{m_1,m_2}}$ and let $\gam_m^{1D}$ denote its spectral gap.

Then, $H_{m}^{1D}$ is frustration-free and there exist positive constants
\beq\label{eq:Cdependence}
C_1^{1D}=C_1^{1D}(m_2,R,\{P^S\}_{S\in\curly{S}}),\qquad C_2^{1D}=C_2^{1D}(m_2,R,\{P^S\}_{S\in\curly{S}}),
\eeq
such that 
\beq\label{eq:propgap}
 C_1^{1D}\gam^{1D}_m\leq \gam_{(m_1,m_2)}\leq C_2^{1D} \gam^{1D}_m.
\eeq
\e{prop}

This proposition allows us to study the spectral gap of the simpler, effective 1D Hamiltonian $H_{m}^{1D}$.

\subsubsection{The one-step coarse-graining procedure}
The reader may find it helpful to consult Figure \ref{fig:1dcg} during the argument. Recall Definition \eqref{eq:HLam0defn} of the Hamiltonian,
$$
H_{\Lam_0}
=\sum_{x\in\Lam_0}\sum_{\substack{S\in\curly{S}:\\ x+S\subset\Lam_0}} P^S_{x+S},
$$
and recall that $\Lam_0=\Lam_{m_1,m_2}$.

Let $1\leq j\leq m$ and $x\in \Lam_j$. We will decompose the index set of interactions $\curly{S}$ into the sets
\beq\label{eq:Sjdefn}
\begin{aligned}
\curly{S}_j(x):=&\setof{S\in\curly{S}}{(x+S)\subset \Lam_{j}},\\
\curly{S}_{j,j+1}(x):=&\setof{S\in\curly{S}}{(x+S)\subset \Lam_j\cup\Lam_{j+1} \;\mathrm{and}\; (x+S)\cap\Lam_{j+1}\neq\emptyset}.
\end{aligned}
\eeq
(We set $\curly{S}_{0,1},\curly{S}_{m,m+1}(x):=\emptyset$.) By construction, $\curly{S}_j(x)$ labels the interactions within a metaspin $\curly{H}_j$ and $\curly{S}_{j,j+1}(x)$ labels the interactions between two neighboring metaspins $\curly{H}_j$ and $\curly{H}_{j+1}$; see Figure \ref{fig:1dcg}.

The key is Assumption \ref{ass:FR}, which says that $\mathrm{diam}(S)<R$ holds for all $S\in\curly{S}$. It implies that, for every $x\in \Lam_j$, we have a complete decomposition
\beq\label{eq:decomp}
\setof{S\in\curly{S}}{x+S\subset \Lam_0}=\curly{S}_{j-1,j}(x)\cup \curly{S}_{j}(x)\cup\curly{S}_{j,j+1}(x).
\eeq
This decomposition allows us to write the Hamiltonian as a 1D FF and \emph{nearest-neighbor} Hamiltonian on metaspins. We define the corresponding interaction terms
$$
\curly{I}_{j}(x):=\sum_{S\in\curly{S}_{j}(x)}P^S_{x+S},\qquad \curly{I}_{j,j+1}(x):=\sum_{S\in\curly{S}_{j,j+1}(x)} P^S_{x+S}.
$$
The decomposition \eqref{eq:decomp} then gives
$$
\begin{aligned}
H_{\Lam_{m_1,m_2}}
=\sum_{j=1}^{m} \sum_{x\in\Lam_j}\l(\curly{I}_{j-1,j}(x)+\curly{I}_{j}(x)+\curly{I}_{j,j+1}(x)\r).
\end{aligned}
$$
We may write this as
\beq\label{eq:Hrewrite}
H_{\Lam_{m_1,m_2}}=\sum_{j=1}^{m-1} \tilde h_{j,j+1}
\eeq
by introducing the following metaspin interactions: For $2\leq j\leq m-2$, let
\beq\label{eq:hjtildedefn}
\begin{aligned}
\tilde h_{j,j+1}:=
\sum\limits_{x\in\Lam_j}\l(\frac{1}{2}\curly{I}_j(x)+\curly{I}_{j,j+1}(x)\r)+\sum\limits_{x\in\Lam_{j+1}}\l(\frac{1}{2}\curly{I}_{j+1}(x)+\curly{I}_{j,j+1}(x)\r)
\end{aligned}
\eeq
and let
\beq\label{eq:hjtildedefn'}
\begin{aligned}
\tilde h_{1,2}:=&\sum\limits_{x\in\Lam_1}\l(\curly{I}_1(x)+\curly{I}_{1,2}(x)\r)+\sum\limits_{x\in\Lam_{2}}\l(\frac{1}{2}\curly{I}_{2}(x)+\curly{I}_{1,2}(x)\r),\\
\tilde h_{m-1,m}:=&\sum\limits_{x\in\Lam_{m-1}}\l(\frac{1}{2}\curly{I}_{m-1}(x)+\curly{I}_{m-1,m}(x)\r)+\sum\limits_{x\in\Lam_{m}}\l(\curly{I}_{m}(x)+\curly{I}_{m-1,m}(x)\r).
\end{aligned}
\eeq
We see that, on metaspins, $H_{\Lam_{m_1,m_2}}$ closely resembles the 1D effective Hamiltonian in Proposition \ref{prop:cg}. To prove the proposition, it remains to replace $\tilde h_{j,j+1}$ by \emph{bona fide} projections $h_{j,j+1}$.

\subsubsection{Proof of Proposition \ref{prop:cg}}
For all $1\leq j\leq m$, we define $\curly{H}_j$ by \eqref{eq:Hjdefn} yielding the Hilbert space decomposition $\curly{H}_{\Lam_{m_1,m_2}}=\bigotimes_{j=1}^m\curly{H}_j$. Note that $\curly{H}_j$ is isomorphic to $\C^{d^{m_2R}}$.

Next we replace $\tilde h_{j,j+1}$ by \emph{bona fide} projections $h_{j,j+1}$ as follows. Given an operator $A$, we write $\Pi_{\ker A}$ for the projection onto $\ker A$. We define the projections
\beq\label{eq:hjdefn}
h_{j,j+1}:=I-\Pi_{\ker \tilde h_{j,j+1}}
\eeq
for every $1\leq j\leq m-1$. It is then immediate that the Hamiltonian
$$
H_{m}^{1D}=\sum_{j=1}^{m-1} h_{j,j+1}
$$
is frustration-free. Indeed,
\beq\label{eq:propff}
\ker H_{m}^{1D}=\ker H_{\Lam_{m_1,m_2}}\neq\{0\}
\eeq
by Assumption \ref{ass:chiralFF}.

We write $\lam_{\min,j}$ ($\lam_{\max,j}$) for the smallest (largest) \emph{positive} eigenvalue of $\tilde h_{j,j+1}$. 

\be{lm}\label{lm:trivial}
We have the operator inequalities $
\lam_{\min,j}h_{j,j+1}\leq \tilde h_{j,j+1}\leq \lam_{\max,j}h_{j,j+1}.$
\e{lm}

\be{proof}
This follows directly from the spectral theorem.
\e{proof}

By combining \eqref{eq:Hrewrite} with Lemma \ref{lm:trivial}, we obtain the operator inequalities
\beq\label{eq:propineq}
\l(\min_{1\leq j\leq m-1}\lam_{\min,j}\r) H_{m}^{1D}\leq H_{\Lam_{m_1,m_2}}\leq \l(\max_{1\leq j\leq m-1}\lam_{\max,j}\r)H_{m}^{1D}.
\eeq

The following lemma is the key result. 

\be{lm}\label{lm:same}
\be{enumerate}[label=(\roman*)]
\item The projections $h_{j,j+1}$ are all defined in terms of the same matrix $P_{\mathrm{eff}}$ on $\C^{d^{m_2R}}\otimes \C^{d^{m_2R}}$, in the sense of \eqref{eq:hjPdefn}. 
\item We have
$$
\begin{aligned}
\lam_{\min,2}&=\lam_{\min,3}=\ldots=\lam_{\min,m-2}=:\lam_{\min},\\
\lam_{\max,2}&=\lam_{\max,3}=\ldots=\lam_{\max,m-2}=:\lam_{\max},\\
\lam_{\min,j}&\geq \lam_{\min},\qquad \lam_{\max,j}\leq 2\lam_{\max},\qquad \textnormal{for }\, j\in \{1,m-1\}.
\end{aligned}
$$
\e{enumerate}
\e{lm}

We first give the 
\be{proof}[Proof of Proposition \ref{prop:cg} assuming Lemma \ref{lm:same}]
It suffices to prove \eqref{eq:propgap}. We apply the estimates from Lemma \ref{lm:same} (ii) to \eqref{eq:propineq} and find
\beq\label{eq:propineq2}
\lam_{\min} H_{m}^{1D}\leq H_{\Lam_{m_1,m_2}}\leq 2\lam_{\max}H_{m}^{1D}.
\eeq
Together, \eqref{eq:propff} and \eqref{eq:propineq2} imply the estimates \eqref{eq:propgap} on the spectral gaps, with the positive constants $C_1$ and $C_2$ given by
$$
C_1^{1D}:=\lam_{\min},\qquad\quad C_2^{1D}:=2\lam_{\max}.
$$
Recall that $\lam_{\min}$ ($\lam_{\max}$) are the smallest (largest) eigenvalues of $\tilde h_{j,j+1}$ for all $2\leq j\leq m-1$, and so they satisfy \eqref{eq:Cdependence}.
\e{proof}

It remains to give the

\be{proof}[Proof of Lemma \ref{lm:same}]\mbox{}\\
\dashuline{Proof of (i).} Notice that all the $\tilde h_{j,j+1}$ defined in \eqref{eq:hjtildedefn} and \eqref{eq:hjtildedefn'} are non-negative and frustration-free, thanks to Assumption \ref{ass:FR} (ii). To see which matrices they correspond to on $\C^{d^{m_2R}}\otimes \C^{d^{m_2R}}$ (in the sense of \eqref{eq:hjPdefn}), we identify $\C^{d^{m_2R}}\otimes \C^{d^{m_2R}}$ with $\curly{H}_1\otimes \curly{H}_2$. Let
$$
\tilde h_{\mathrm{eff}}:=\sum\limits_{x\in\Lam_1}\l(\frac{1}{2}\curly{I}_1(x)+\curly{I}_{1,2}(x)\r)+\sum\limits_{x\in\Lam_{2}}\l(\frac{1}{2}\curly{I}_{2}(x)+\curly{I}_{1,2}(x)\r).
$$
Then, we have the correspondences
\beq\label{eq:correspondence}
\tilde h_{j,j+1}\rightarrow \be{cases}
\tilde h_{\mathrm{eff}}, &\textnormal{ for }2\leq j\leq m-2,\\
\tilde h_{\mathrm{eff}}+\frac{1}{2}\sum\limits_{x\in\Lam_1} \curly{I}_1(x), &\textnormal{ for }j=1,\\
\tilde h_{\mathrm{eff}}+\frac{1}{2}\sum\limits_{x\in\Lam_2}\curly{I}_2(x), &\textnormal{ for }j=m-1.
\e{cases}
\eeq
where ``$\rightarrow$'' refers to correspondence in the sense of \eqref{eq:hjPdefn} (there, $h_{i,i+1}\rightarrow P$). In particular, $X\rightarrow Y$ implies that $X$ has the same spectrum as $Y$ up to a trivial degeneracy.

Since all the matrices on the right-hand side of \eqref{eq:correspondence} have the same kernel, the definition \eqref{eq:hjdefn} implies that, for all $2\leq j\leq m-1$,
\beq\label{eq:ker}
h_{j,j+1}=I-\Pi_{\ker \tilde h_{j,j+1}}\rightarrow  I-\Pi_{\ker \tilde h_{\mathrm{eff}}}=:P_{\mathrm{eff}}.
\eeq
This proves statement (i) of Lemma \ref{lm:same}.\\

\dashuline{Proof of (ii).} Since the correspondence $X\rightarrow Y$ implies equality of the spectra up to a trivial degeneracy, we have $\lam_{\min}=\lam_{\min,j}$ and $\lam_{\max}=\lam_{\max,j}$ for all $2\leq j\leq m-2$. Moreover, we have the operator inequality
$$
\tilde h_{\mathrm{eff}}\leq \tilde h_{\mathrm{eff}}+\frac{1}{2}\sum\limits_{x\in\Lam_1} \curly{I}_1(x)\leq 2\tilde h_{\mathrm{eff}}.
$$
In the same way as above, using also the equality of kernels \eqref{eq:ker}, we conclude the eigenvalue inequalities in statement (ii). This concludes the proof of Lemma \ref{lm:same} and hence of Proposition \ref{prop:cg}.
\e{proof}

\subsection{Step 2: Reduction to ``macroscopic'' gaps}\label{sect:macro}
Let $m_1$ and $m_2$ be positive integers. Without loss of generality, we assume that the finite interaction range $R>0$, which is guaranteed to exist by Assumption \ref{ass:FR} (i), divides $m_1$, i.e., $m_1= m R$ for some integer $m$. (If this is not initially the case, we can replace $R$ with $\ceil{R/m_1}m_1\geq R$.)

Let $4\leq n\leq m/2$. By Proposition \ref{prop:cg} and Theorem \ref{thm:main2}, applied to the effective Hamiltonian $H_m^{1D}$, we find
\beq\label{eq:firststep}
\begin{aligned}
&\gam_{(m_1,m_2)}\geq C_1^{1D}\gam_m^{1D}\geq C_n\l(\min\l\{\gam_n^{1D},\min_{0\leq j\leq n-2}\frac{\sum_{k=j}^{n-2} c_{k-j} \gam^{1D}_{k+1}}{\sum_{k=j}^{n-2} c_{k-j} }\r\}-2\sqrt{6}n^{-3/2}\r),
\end{aligned}
\eeq
where we introduced $C_n:=(2^8\sqrt{6n})^{-1}C_1^{1D}$. We also used that all notions of spectral gap agree when $\Pi_1=\Pi_m=0$.\\

Now we can apply Proposition \ref{prop:cg} again, to the right-hand side in \eqref{eq:firststep} to obtain an inequality that \emph{only involves the spectral gaps of} $H_{\Lam_0}$. For this application of Proposition \ref{prop:cg}, we replace $m_1$ by $m_1'=n R$, respectively $m_1''=(k+1)R$ (and we take the same $m_2$ as before). We get
\beq\label{eq:secondstep}
\begin{aligned}
\gam_{(m_1,m_2)}\geq &C_n (C_2^{1D})^{-1}\l(\min\l\{\gam_{(nR,m_2)},\min_{0\leq j\leq n-2}\frac{\sum_{k=j}^{n-2} c_{k-j} \gam_{((k+1)R,m_2)}}{\sum_{k=j}^{n-2} c_{k-j} }\r\}-C_2^{1D} \frac{2\sqrt{6}}{n^{3/2}}\r),
\end{aligned}
\eeq

This already looks similar to the claimed inequality \eqref{eq:thmchiral}. The final (and crucial) ingredient is to control the weighted average in \eqref{eq:secondstep} in terms of ``macroscopic'' gaps only. We estimate
$$
\gam_{((k+1)R,m_2)}\geq\be{cases}
\min\limits_{\floor{n/2}\leq l\leq n-1} \gam_{(lR,m_2)},&\mathrm{if}\;\floor{n/2}-1\leq k\leq n-2,\\
\;\;0,&\mathrm{if}\;1\leq k\leq \floor{n/2}-2.
\e{cases}
$$
Employing these estimates on the fraction in \eqref{eq:secondstep} gives
$$
\begin{aligned}
&\min_{0\leq j\leq n-2}
\l(\sum_{k=j}^{n-2} c_{k-j} \gam_{(k+1)R,m_2}
\bigg/\sum_{k=j}^{n-2} c_{k-j} \r)\\
\geq
&\l(\min_{0\leq j\leq n-2}\sum_{k=\max\{j,\floor{n/2}-1\}}^{n-2} c_{k-j}  
\bigg/\sum_{k=j}^{n-2} c_{k-j} \r) \min\limits_{\floor{n/2}\leq l\leq n-1} \gam_{(lR,m_2)}.
\end{aligned}
$$
For $\floor{n/2}-1\leq j\leq n-2$, the ratio of the two sums is equal to $1$. 

Let $0\leq j\leq \floor{n/2}-2$. By Assumption \ref{ass:c}, we have $c_{j+1}\geq c_{j}$ and $c_j=c_{n-2-j}$, hence
$$
\l(\sum_{k=n/2-1}^{n-2} c_{k-j}  
\bigg/\sum_{k=j}^{n-2} c_{k-j} \r)
\geq \l(\sum_{k=0}^{\floor{n/2}-1} c_{k}  
\bigg/\sum_{k=0}^{n-2} c_{k} \r)\geq \frac{1}{2}.
$$
Applying these estimates to \eqref{eq:secondstep} yields the claimed inequality \eqref{eq:thmchiral} with the constants
$$
C_1:=\frac{C_n}{2C_2^{1D}}=\frac{C_1^{1D}}{2^9C_2^{1D}\sqrt{6n} },\qquad
C_2:=C_2^{1D} 4\sqrt{6}.
$$
Here we used that $C_n=(2^8\sqrt{6n})^{-1}C_1^{1D}$. From the dependencies \eqref{eq:Cdependence} of $C_1^{1D}$ and $C_2^{1D}$, we can read off \eqref{eq:cdependence} and this finishes the proof Theorem \ref{thm:chiral}.
\qed

\section{Proof of the 2D finite-size criterion}
The proof of Theorem \ref{thm:2D} follows the general line of argumentation for Theorems \ref{thm:main1} and \ref{thm:main2} -- but now in two dimensions. The main complications arise from the richer geometry of 2D shapes and it is essential for our argument that the local patches we consider are rhomboidal.

\subsection{Proof strategy}
In step 1 of the proof we perform a one-step coarse-graining procedure to replace $H_{\curly{R}_{m_1,m_2}}$ by an effective nearest-neighbor Hamiltonian on ``plaquette metaspins''. This is a genuinely 2D analog of Proposition \ref{prop:cg} and may be of independent interest.

The basic idea of the one-step coarse-graining procedure is that all the spins in a $Q(y)$-box (i.e., a box in Figure \ref{fig:rhomboidalpatch}) are replaced by a single metaspin. The effective 2D Hamiltonian is then defined on plaquette metaspins and is nearest-neighbor, in the sense that only plaquettes that touch can interact (this holds because the plaquettes arise from boxes whose sidelength is the interaction range $R$).\\

In step 2 of the proof, we implement a Knabe-type argument in 2D inspired by \cite{GM}. Namely, we construct the Hamiltonian out of deformed patch operators. It is essential that our patches are rhomboidal: a certain combinatorial property of them is the key for Proposition \ref{prop:2D} (see Remark \ref{rmk:W}). Another geometrical property of rhomboids is used to get Lemma \ref{lm:n1n2}: Their intersections are still rhomboids; see Figure \ref{fig:intersect}.

In step 3 of the proof, we reduce the relevant spectral gaps to ``macroscopic'' ones, i.e., ones whose linear size is comparable to $n$. This is a generalization of the argument in Section \ref{sect:macro} to 2D.

In step 4, we derive the $\Theta(n^{-3/2})$  local gap threshold by choosing the appropriate deformation parameters defining the deformed patch operators. 

\subsection{Step 1: The effective 2D Hamiltonian}
\begin{figure}[t]
\begin{center}
\includegraphics[scale=.48]{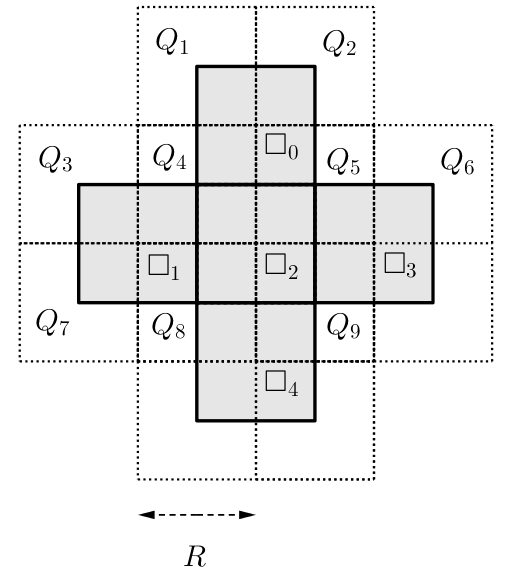}
\end{center}
\caption{The shaded set is $\curly{D}_{2,2}$ -- formally defined in \eqref{eq:Ddefn}. It is constructed from the dotted set $\curly{R}_{2,2}$ as follows: $\curly{D}_{2,2}$ is the set of plaquettes (= sites of the dual lattice) which touch exactly four sites in $\curly{R}_{2,2}$. The key fact is that only boxes (now plaquette metasites) $Q(x)$ that touch can interact. (This makes the effective 2D Hamiltonian nearest-neighbor.) For coarse-graining, we write the original Hamiltonian as $H_{\curly{R}_{m_1,m_2}}=\sum_{\Box\in\curly{D}_{m_1,m_2}} \tilde h_{\Box}$ where each $\tilde h_{\Box}$ is obtained by equidistributing the interaction terms in $H_{\curly{R}_{m_1,m_2}}$ among all eligible plaquettes.}
\label{fig:plaq}
\end{figure}

We now present the details of the one-step coarse-graining procedure. We recommend to skip these on a first reading and to focus on Proposition \ref{prop:2Dcg} and Figure \ref{fig:plaq}.

We recall Definition \ref{defn:rhomb} of a rhomboidal patch,
\beq\label{eq:rhombrecall}
\curly{R}_{m_1,m_2}=\bigcup_{j=0}^{m_1-1}\bigcup_{j'=0}^{m_2} Q(j f_1+j'f_2)
\cup \bigcup_{j=0}^{m_1}\bigcup_{j'=0}^{m_2-1} Q(Re_2+j f_1+j'f_2),
\eeq
which was depicted in Figure \ref{fig:rhomboidalpatch}. 

Recall that we assume that $R$ is an odd integer. We group all the spins within a box $Q(y)$ (i.e., any box in Figure \ref{fig:rhomboidalpatch}) into a single metaspin. This amounts to the Hilbert space decomposition
$$
\begin{aligned}
\curly{H}_{\curly{R}_{m_1,m_2}}=&\bigotimes_{y\in \curly{L}_{m_1,m_2}} \curly{H}_y\\
\curly{L}_{m_1,m_2}:=&\bigcup_{j=0}^{m_1-1}\bigcup_{j'=0}^{m_2} \{j f_1+j' f_2\}
\cup \bigcup_{j=0}^{m_1}\bigcup_{j'=0}^{m_2-1} \{Re_2+j f_1+j' f_2\}
\end{aligned}
$$
where each $\curly{H}_y$ is isomorphic to $\C^{d|Q(0)|}$. The set $\curly{L}_{m_1,m_2}\subset \curly{R}_{m_1,m_2}$ labels the metaspins. (It consists of the centers $y$ of all the boxes $Q(y)$ in \eqref{eq:rhombrecall}.)

Since the boxes have sidelength equal to the interaction range $R$, only boxes that touch (either along a side or a corner) can interact in $H_{\curly{R}_{m_1,m_2}}$. Equivalently, only metaspins belonging to the same plaquette can interact in the effective 2D Hamiltonian. To formalize this, we introduce the set $\curly{D}_{m_1,m_2}$ of \emph{plaquettes} on the $\curly{L}_{m_1,m_2}$ lattice that \emph{touch four sites} of $\curly{L}_{m_1,m_2}$. ($\curly{D}_{m_1,m_2}$ is a finite subset of the dual lattice to $\curly{L}_{m_1,m_2}$.) The set $\curly{D}_{2,2}$ (corresponding to $\curly{R}_{2,2}$ from Figure \ref{fig:rhomboidalpatch}) is depicted in Figure \ref{fig:plaq}.

Formally, 
\beq\label{eq:Ddefn}
\curly{D}_{m_1,m_2}:=\bigcup_{j=0}^{m_1-1}\bigcup_{j'=0}^{m_2-1} \{j f_1+(j'+1/2)f_2\}
\cup \bigcup_{j=0}^{m_1-2}\bigcup_{j'=0}^{m_2-2} \{Re_1+j f_1+(j'+1/2)f_2\}.
\eeq
(Note the shift by $f_2/2$, it amounts to changing to the dual lattice.) Given a site $y\in\curly{L}_{m_1,m_2}$ and a plaquette $\Box\in\curly{D}_{m_1,m_2}$, we write $y\in \Box$ if $y$ is one of the four vertices of $\Box$, otherwise we write $y\notin \Box$.

The upshot of the one-step coarse-graining procedure is the following 2D analog of Proposition \ref{prop:cg}.

\be{prop}[Effective 2D Hamiltonian]\label{prop:2Dcg}
Let $m_1$ and $m_2$ be positive integers and let $R$ be an odd integer. There exists a projection $P_{\mathrm{eff}}$ acting on $\l(\C^{d|Q(0)|}\r)^{\otimes 4}$, such that the following holds. For each plaquette $\Box\in \curly{D}_{m_1,m_2}$, we define a projection on $\curly{H}_{\curly{R}_{m_1,m_2}}$ via
\beq\label{eq:hBoxdefn}
h_\Box:= P_{\mathrm{eff}}\otimes \bigotimes_{\substack{y\in \curly{L}_{m_1,m_2}\\ y\not\in \Box}} I_y.
\eeq
We define the Hamiltonian
\beq\label{eq:H2Ddefn}
H^{2D}_{m_1,m_2}:=\sum_{\Box\in \curly{D}_{m_1,m_2}} h_\Box
\eeq
on $\curly{H}_{\curly{R}_{m_1,m_2}}$ and write $\gam^{2D}_{m_1,m_2}$ for its spectral gap.

Then, $H^{2D}_{m_1,m_2}$ is frustration-free and there exist positive constants
 \beq\label{eq:2DCdependence}
C_1^{2D}=C_1^{2D}(R,\{P^S\}_{S\in\curly{S}}),\qquad C_2^{2D}=C_2^{2D}(R,\{P^S\}_{S\in\curly{S}})
\eeq
such that 
\beq\label{eq:prop2Dgap}
C_1^{2D}\gam^{2D}_{m_1,m_2}\leq \gam_{\curly{R}_{m_1,m_2}}\leq C_2^{2D} \gam^{2D}_{m_1,m_2}.
\eeq
\e{prop}

The argument is a 2D generalization of the proof of Proposition \ref{prop:cg}. In part A, we rewrite the \emph{original} Hamiltonian $H_{\curly{R}_{m_1,m_2}}$ in terms of plaquettes; see \eqref{eq:plus} below. In part B, we define the \emph{effective} Hamiltonian $H^{2D}_{m_1,m_2}$ by replacing the plaquette operators (called $\tilde h_\Box$) by projections onto their range (see \eqref{eq:hboxdefn}) --- these projections are then the operators $h_\Box$ appearing above. Since the spectral gaps of $\tilde h_\Box$ and $h_\Box$ are related by universal factors $C_1^{2D},C_2^{2D}$ (which are different at the bulk and at the edge, just as in the 1D case), we obtain the claimed gap relation \eqref{eq:prop2Dgap}.

\subsubsection{Part A: Organizing the interactions into plaquettes}
In part A, we rewrite the original Hamiltonian $H_{\curly{R}_{m_1,m_2}}$ in terms of coarse-grained metasites. The effective Hamiltonian $H^{2D}_{m_1,m_2}$ only appears in part B.

We recommend the reader consider Figure \ref{fig:plaq}. The boxes $\{Q(y)\}_{y\in \curly{L}_{m_1,m_2}}$ take the role of the rectangles $\{\Lam_j\}_{1\leq j\leq m-1}$.  

We first decompose the Hamiltonian in terms of metasites $y\in \curly{L}_{m_1,m_2}$, i.e.,
$$
H_{\curly{R}_{m_1,m_2}}=\sum_{x\in \curly{R}_{m_1,m_2}}\sum_{\substack{S\in\curly{S}:\\ x+S\subset\Lam_0}} P^S_{x+S}
=\sum_{y\in \curly{L}_{m_1,m_2}}\sum_{x\in Q(y)}\sum_{\substack{S\in\curly{S}:\\ x+S\subset\Lam_0}} P^S_{x+S}.
$$
Since the boxes $Q(y)$ in \eqref{eq:rhombrecall} have sidelength equal to the interaction range $R$ (in standard graph distance), only boxes that touch can interact. We can thus group the interaction terms $P^S_{x+S}$ for every $x\in\curly{R}_{m_1,m_2}$, similarly to \eqref{eq:Sjdefn}. There are only three kinds of interaction terms.
\be{itemize}
\item[(a)] Interactions within a single box $Q(y)$.
$$
\curly{I}_{Q(y)}:=\sum_{x\in Q(y)}\sum_{\substack{S\in\curly{S}:\\ x+S\subset Q(y)}} P^S_{x+S}.
$$ 
\item[(b)] Interactions between two boxes $Q(y_1)$ and $Q(y_2)$ that share a side.
$$
\curly{I}_{Q(y_1),Q(y_2)}:=\sum_{x\in Q(y_1)\cup Q(y_2)}\sum_{\substack{S\in\curly{S}:\\ 
x+S\subset Q(y_1)\cup Q(y_2)\\
(x+S)\cap Q(y_1)\neq \emptyset\\
(x+S)\cap Q(y_2)\neq \emptyset
}} P^S_{x+S}.
$$
\item[(c)] Interactions between four boxes $Q(y_1),Q(y_2),Q(y_3)$ and $Q(y_4)$ that share a corner.
$$
\curly{I}_{Q(y_1),Q(y_2),Q(y_3),Q(y_4)}:=\sum_{x\in \bigcup_{i=1}^4 Q(y_i)}\sum_{\substack{S\in\curly{S}:\\ \forall 1\leq i\leq 4:\\ (x+S)\cap Q(y_i)\neq \emptyset}} P^S_{x+S}.
$$
\e{itemize}

Given that only boxes that touch can interact because of the interaction range, \emph{notable omissions} from this list are interaction that involve (d) exactly two boxes that touch only at a corner or (e) exactly three boxes. The absence of such interactions is a consequence of Assumption \ref{ass:convex} and the fact that none of the admissible interaction shapes described there (i.e., lines, or boxes of diameter $\geq 2$) can intersect two boxes that touch only at a corner, without already involving the other two boxes that touch this corner. The absence of interaction types (d) and (e) will be used in the following.\\
 
We now rewrite the original Hamiltonian in terms of plaquettes, i.e.,
\beq\label{eq:plus}
 H_{\curly{R}_{m_1,m_2}}=\sum_{\Box\in\curly{D}_{m_1,m_2}} \tilde h_{\Box},
\eeq
where each $\tilde h_\Box$ acts non-trivially only on metasites $y\in \Box$ as in \eqref{eq:hBoxdefn}.

We first give the definition of the plaquette operator $\tilde h_\Box$ in words, generalizing the ``equidistribution rule'' used in the proof of Proposition \ref{prop:cg}: 

\emph{Each $\tilde h_\Box$ is defined by equidistributing an interaction term among those plaquettes in $\curly{D}_{m_1,m_2}$ that touch all the boxes which are involved in that term.} (Recall that there are three basic types of interaction terms which are classified by (a)-(c) above.)

Let us give a more formal definition. Let $\Box\in\curly{D}_{m_1,m_2}$ be arbitrary. By definition of the sets $\curly{D}_{m_1,m_2}$ and $\curly{R}_{m_1,m_2}$ (cf.\ Figure \ref{fig:plaq}),  there exist four boxes $Q_i,Q_j,Q_k,Q_l\in \curly{R}_{m_1,m_2}$ that touch $\Box$. We suppose that these boxes are labeled in clockwise order. The general form of $\tilde h_\Box$ is then
\beq\label{eq:tildehboxdefn}
\tilde h_\Box:=p_{i}\curly{I}_{i}+p_{j}\curly{I}_{j}+p_{k}\curly{I}_{k}+p_{l}\curly{I}_{l}
+p_{i,j}\curly{I}_{i,j}+p_{j,k}\curly{I}_{j,k}+p_{k,l}\curly{I}_{k,l}+p_{l,i}\curly{I}_{l,i},
+\curly{I}_{i,j,k,l}
\eeq
where $i,j,\ldots$ labels the box $Q_i,Q_j,\ldots$ and the coefficients $p_i,p_{i,j},\ldots \in \left\{\frac{1}{4},\frac{1}{3},\frac{1}{2},1\right\}$ are determined by the equidistribution procedure described above. In particular, the interaction term between all four boxes $Q_i,Q_j,Q_k,Q_l$, i.e., the type (c) term $\curly{I}_{i,j,k,l}$, is fully assigned to the unique plaquette which touches all the four boxes. Hence, we always have $p_{i,j,k,l}=1$, as can be seen in \eqref{eq:tildehboxdefn}.  

We emphasize that in the Definitions \eqref{eq:plus},\eqref{eq:tildehboxdefn}, we implicitly used Assumption \ref{ass:convex} (i), specifically that all the interaction terms are classified by (a)-(c) given above and that there are no diagonal $2$-box and no $3$-box interactions, called (d) and (e) above. Indeed, the absence of such interactions implies that the boundary corners do not require their own plaquette. For example, all the interactions involving the boxes $Q_1,Q_3,Q_4$ in Figure \ref{fig:plaq} are contained in the plaquettes $\Box_0,\Box_1,\Box_2$ --- precisely because there are no interactions that involve the two boxes $Q_1$ and $Q_3$ simultaneously.\\

 To consolidate the Definition \eqref{eq:tildehboxdefn} of $\tilde h_\Box$, let us give some concrete examples. In Figure \ref{fig:plaq}, the plaquettes $\Box_0,\Box_2$ are assigned the following interaction terms:
\beq\label{eq:hboxdefn1}
\begin{aligned}
\tilde h_{\Box_0}=&\curly{I}_{1}+\curly{I}_{2}+\frac{1}{3}(\curly{I}_{4}+\curly{I}_{5})
+\curly{I}_{1,2}+\curly{I}_{1,4}+\curly{I}_{2,5}+\frac{1}{2}\curly{I}_{4,5}
+\curly{I}_{1,2,4,5},\\
\tilde h_{\Box_2}=&\frac{1}{3}(\curly{I}_{4}+\curly{I}_{5}+\curly{I}_{8}+\curly{I}_{9})
+\frac{1}{2}(\curly{I}_{4,5}+\curly{I}_{5,9}+\curly{I}_{4,8}+\curly{I}_{8,9})
+\curly{I}_{4,5,8,9}.
\end{aligned}
\eeq

Let us give another example. We say a plaquette in $\curly{D}_{m_1,m_2}$ lies in the ``bulk'' of the system, if all of the boxes that it touches (including at a corner) are at Euclidean distance $\geq 2$ from the complement of $\curly{R}_{m_1,m_2}$. Otherwise, the plaquette lies at the ``edge'', and we denote the set of edge plaquettes by $\del\curly{D}_{m_1,m_2}$. According to this definition, all of the plaquettes in Figure \ref{fig:plaq} are ``edge'' plaquettes. Let us also write down $\tilde h_\Box$ for a bulk plaquette $\Box_B\in \curly{D}_{m_1,m_2}\setminus\del\curly{D}_{m_1,m_2}$ :
\beq\label{eq:hboxdefn2}
\tilde h_{\Box_B}=\frac{1}{4}(\curly{I}_{i}+\curly{I}_{j}+\curly{I}_{k}+\curly{I}_{l})
+\frac{1}{2}(\curly{I}_{i,j}+\curly{I}_{j,k}+\curly{I}_{k,l}+\curly{I}_{l,i})
+\curly{I}_{i,j,k,l}.
\eeq
Notice that the coefficients $p_i,p_{i,j},\ldots$ appearing in \eqref{eq:hboxdefn2} are less or equal to those appearing in \eqref{eq:hboxdefn1}. The reason is that interaction terms in the bulk have to be equidistributed among the largest possible number of plaquettes. (E.g., each bulk box $Q_i$ touches four bulk plaquettes, so the equidistribution rule leads to the prefactor $\frac{1}{4}$ in $\tilde h_{\Box_B}$.)

This establishes formula \eqref{eq:plus} which expresses the original Hamiltonian in terms of plaquettes.

\be{rmk}
As in the 1D case, compare \eqref{eq:hjtildedefn} and \eqref{eq:hjtildedefn'}, we see that $\tilde h_\Box$ depends on the position of the plaquette relative to the edge. Nonetheless, all the operators $\tilde h_\Box$ contain the same interaction terms modulo translations and up to a different prefactor, and therefore the $\tilde h_\Box$ have isomorphic kernels. This observation is at the core of our argument; see Lemma \ref{lm:same2D} below.
\e{rmk}

\subsubsection{Part B: Defining the effective Hamiltonian and proof of Proposition \ref{prop:2Dcg}}
We now replace the plaquette operators $\tilde h_\Box$ by projections to define the effective Hamiltonian. Analogous to \eqref{eq:hjdefn}, for every $\Box\in\curly{D}_{m_1,m_2}$, we define the projection
\beq\label{eq:hboxdefn}
h_\Box:=I-\Pi_{\ker \tilde h_\Box}.
\eeq
It is then immediate that the effective Hamiltonian
$$
H^{2D}_{m_1,m_2}:=\sum_{\Box\in \curly{D}_{m_1,m_2}} h_\Box
$$
is frustration-free. Indeed,
\beq\label{eq:2DFF}
\ker H^{2D}_{m_1,m_2}=\ker H_{\curly{R}_{m_1,m_2}}\neq \{0\}
\eeq
by Assumption \ref{ass:convex} (ii).

The following analog of Lemma \ref{lm:trivial} follows directly from the spectral theorem. Let $\lam_{\min,\Box}$ ($\lam_{\max,\Box}$) denote the smallest (largest) \emph{positive} eigenvalue of $\tilde h_\Box$.

\be{lm}
We have the operator inequalities
$
\lam_{\min,\Box}h_\Box\leq \tilde h_\Box\leq \lam_{\max,\Box}h_\Box.
$
\e{lm}

This lemma gives
\beq\label{eq:2Dprelim}
\l(\min_{\Box\in \curly{D}_{m_1,m_2}}\lam_{\min,\Box}\r)H^{2D}_{m_1,m_2}\leq H_{\curly{R}_{m_1,m_2}}\leq \l(\max_{\Box\in \curly{D}_{m_1,m_2}}\lam_{\max,\Box}\r)H^{2D}_{m_1,m_2}.
\eeq

Recall that $\del\curly{D}_{m_1,m_2}$ denotes the edge plaquettes. The following lemma is a genuinely 2D analog of Lemma \ref{lm:same}. 

\be{lm}\label{lm:same2D}
\be{enumerate}[label=(\roman*)]
\item The projections $\{h_\Box\}_{\Box\in\curly{D}_{m_1,m_2}}$ are defined in terms of the same matrix $P_{\mathrm{eff}}$ on $\l(\C^{d|Q(0)|}\r)^{\otimes 4}$.
\item The bulk operators $\{\tilde h_{\Box_B}\}_{\Box_B\in\curly{D}_{m_1,m_2}}\setminus \del \curly{D}_{m_1,m_2}$ are isospectral. We write $\lam_{\min}$ ($\lam_{\max}$) for their smallest (largest) positive eigenvalue.

\item We have
\beq\label{eq:2Dkey}
\lam_{\min}\leq\min_{\Box\in \curly{D}_{m_1,m_2}}\lam_{\min,\Box},\qquad \max_{\Box\in \curly{D}_{m_1,m_2}}\lam_{\max,\Box}\leq 4\lam_{\max}.
\eeq
\e{enumerate}
\e{lm}

\be{proof}[Proof of Lemma \ref{lm:same2D}]
The statement essentially holds by construction of the operators $\tilde h_\Box$; recall Definition \eqref{eq:tildehboxdefn} of $\tilde h_\Box$, the examples in \eqref{eq:hboxdefn1} and \eqref{eq:hboxdefn2}, and  the remarks following the latter equation. 

Our original 2D Hamiltonian $H_{\curly{R}_{m_1,m_2}}$ is assumed to be translation-invari\-ant in the bulk, in the sense that its interactions are defined by shifting a fixed ``unit cell of interactions'' around the microscopic lattice $\curly{R}_{m_1,m_2}$ (see \eqref{eq:HRdefn}). In particular, the local interactions $P^S_{x+S}$ and $P^S_{y+S}$, with $x,y\in \curly{R}_{m_1,m_2}$, are unitarily equivalent by translation. Hence, the various interaction terms in the classification (a)-(c) are unitarily equivalent, if the corresponding boxes are related by a translation. For instance, $\curly{I}_i$ is unitarily equivalent to $\curly{I}_{i'}$ for any two boxes $Q_{i},Q_{i'}\in \curly{R}_{m_1,m_2}$ and, similarly, $\curly{I}_{i,j}$ is unitarily equivalent to $\curly{I}_{i',j'}$ if the pairs of boxes $(Q_i,Q_j)$ and $(Q_{i'},Q_{j'})$ are themselves related by a translation.

The unitary equivalence of the individual interaction terms already proves statements (i) and (ii). Indeed, for (i), we note that the operator $h_\Box$ projects onto the range of $\tilde h_\Box$, which is independent of the coefficients $p_i,p_{i,j},\ldots$ (since $\tilde h_\Box$ is a frustration-free sum of non-negative operators). For (ii), we note that formula \eqref{eq:hboxdefn2} holds for all bulk plaquettes $\Box_B\in\curly{D}_{m_1,m_2}\setminus \del \curly{D}_{m_1,m_2}$, for the appropriate boxes $Q_i,Q_j,Q_k,Q_l$. That is, the coefficients are $p_i=p_j=\ldots=\frac{1}{4}$ and $p_{i,j}=p_{j,k}=\ldots=\frac{1}{2}$ for all bulk plaquettes and so (ii) follows by the unitary equivalence of the individual interaction terms.

For statement (iii), recall that the coefficients $p_i,p_{i,j},\ldots \in \left\{\frac{1}{4},\frac{1}{3},\frac{1}{2},1\right\}$ in \eqref{eq:tildehboxdefn} are determined by equidistributing the corresponding interaction terms among all available plaquettes, and are therefore minimal in formula \eqref{eq:hboxdefn2} for the bulk plaquettes $\Box_B$, where they take the values $p_i=p_j=\ldots=\frac{1}{4}$ and $p_{i,j}=p_{j,k}=\ldots=\frac{1}{2}$. The minimality directly gives the first inequality in \eqref{eq:2Dkey}, and the second inequality follows from the fact that
$$
4\tilde h_{\Box_B}:=\curly{I}_{i}+\curly{I}_{j}+\curly{I}_{k}+\curly{I}_{l}
+2(\curly{I}_{i,j}+\curly{I}_{j,k}+\curly{I}_{k,l}+\curly{I}_{l,i})
+4\curly{I}_{i,j,k,l}
$$
has all of its coefficients $\geq 1$. This proves (iii) and hence Lemma \ref{lm:same2D}.
\e{proof}

We are now ready to give the

\be{proof}[Proof of Proposition \ref{prop:2Dcg}]
We apply \eqref{eq:2Dkey} to \eqref{eq:2Dprelim} and find
$$
\lam_{\min}H^{2D}_{m_1,m_2}\leq H_{\curly{R}_{m_1,m_2}}\leq 4\lam_{\max}H^{2D}_{m_1,m_2}.
$$
Together with \eqref{eq:2DFF}, this yields Proposition \ref{prop:2Dcg} with the constants $C_1^{2D}=\lam_{\min}$ and $C_2^{2D}=4\lam_{\max}$. \e{proof}

\subsection{Deformed rhomboidal patch operators}
Thanks to Proposition \ref{prop:2Dcg}, specifically \eqref{eq:prop2Dgap}, it suffices to prove Theorem \ref{thm:2D} for the effective 2D Hamiltonian
$$
H^{2D}_{m_1,m_2}=\sum_{\Box\in \curly{D}_{m_1,m_2}} h_\Box.
$$
We employ a Knabe-type argument in 2D, constructing the Hamiltonian $H^{2D}_{m_1,m_2}$ from (deformed) rhomboidal patch operators; see Definition \eqref{eq:2DBdefn} below. Essentially, the intervals we considered in 1D are replaced by rhomboids in 2D (i.e., rectangles in the $\ell^1$ distance).\\

From now on, we let $n$ be an \emph{even} integer.

As in 1D, the deformed patch operators are defined in terms of a set of coefficients $\{c(1),\ldots,c\l(\frac{n}{2}\r)\}$, on which we make 

\be{ass}\label{ass:c2D}
For every even integer $n\geq 2$, the positive coefficients $\{c(1),\ldots,c\l(\frac{n}{2}\r)\}$ form a non-increasing sequence.
\e{ass}

We write $\curly{P}$ for the set of all plaquettes on $\Z^2$. (It is of course isomorphic to $\Z^2$.) 

\be{defn}[Deformed patch operators]\label{defn:patchdefn}

Let $m_1,m_2$ be positive integers. Fix a plaquette $\boxtimes\in \curly{P}$. 

\be{enumerate}[label=(\roman*)]
\item Let $n\geq 2$ be an even integer. We write $\curly{D}_{n_1,n_2}(\boxtimes)$ for the rhomboid $\curly{D}_{n_1,n_2}$ centered at the plaquette $\boxtimes$. Then we define the patch
\beq\label{eq:intersect}
P_{n,\boxtimes}:=\curly{D}_{n,n}(\boxtimes)\cap \curly{D}_{m_1,m_2}.
\eeq

\item Let $n\geq 2$ be an even integer. We define a distance function $d_{\boxtimes}$ on each $P_{n,\boxtimes}$ via
\beq\label{eq:dboxdefn}
d_{\boxtimes}(\Box):=\min\setof{1\leq k\leq n/2}{\Box\in P_{2k,\boxtimes}}, 
\eeq
for all $\Box\in P_{n,\boxtimes}$.

\item The deformed patch operators are defined by
\beq\label{eq:2DBdefn}
B_{n,\boxtimes} :=\sum_{\Box\in P_{n,\boxtimes}} c(d_\boxtimes(\Box)) h_\Box,
\eeq
where the coefficients $\{c(1),\ldots,c\l(\frac{n}{2}\r)\}$ satisfy Assumption \ref{ass:c2D}.
\e{enumerate}
\e{defn}

Note that $d_{\boxtimes}$ implicitly depends on $n$, but in a trivial way. Note also that $d_{\boxtimes}(\boxtimes)=1$.

\subsection{Step 2: Knabe-type argument in 2D}
 The following 2D analog of Proposition \ref{prop:rewrite} is at the heart of the argument.

The goal is to compute and bound $(H^{2D}_{m_1,m_2})^2$. It is clear that in the expression for $(H^{2D}_{m_1,m_2})^2$, certain terms are present only in the bulk, and not at the edge. A convenient way of bookkeeping for us is as follows: In Proposition \ref{prop:2D} below, we compute and bound $(H^{2D}_{m_1,m_2})^2$ by extending the Hamiltonian to \emph{zero} on all of $\curly{P}$ ($=$ the set of all plaquettes in $\Z^2$). This helps because when we temporarily forget which terms are zero, there is no more edge. (Of course, eventually, we have to recall which terms actually appear and this is spelled out in Lemma \ref{lm:n1n2} later.)

This step is close in spirit to the proof of Theorem 5 in \cite{GM}. However, it is important that we use rhomboids here.

\be{prop}[Knabe-type bound in 2D]\label{prop:2D}
Let $n\geq 2$ be an even integer and let $\{c(1),\ldots, c(n/2)\}$ satisfy Assumption \ref{ass:c2D}. Let
\beq\label{eq:albetadefn}
\begin{aligned}
\al_n:=&\l(2c(1)^2+\sum_{r=2}^{n/2} \l(c(r)^2 (8r-6) +c(r)c(r-1)(8r-10)\r)\r)^{-1},\\
 \beta_n:=&\al_n \l(5c(1)^2+\sum_{r=2}^{n/2} c(r)^2 (16r-12)\r)-1.
\end{aligned}
\eeq
Then
\beq\label{eq:prop2D}
\l(H^{2D}_{m_1,m_2}\r)^2 \geq \al_n \sum_{\boxtimes \in\curly{P}} B_{n,\boxtimes}^2 -\beta_n H^{2D}_{m_1,m_2}.
\eeq
\e{prop}

Recall Definition \ref{defn:patchdefn}. From it, we see that $B_{n,\boxtimes}$ only acts on plaquettes which are at graph distance at most $n/2$ from the center plaquette $\boxtimes$ \emph{and} which lie in the large rhomboid $\curly{D}_{m_1,m_2}$ (cf.\ \eqref{eq:intersect}). Therefore, the sum over $\boxtimes \in \curly{P}$ on the right-hand side of \eqref{eq:prop2D} is actually only taken over a finite set, namely over an $n/2$-collar around the large rhomboid $\curly{D}_{m_1,m_2}$. Adding in the additional zero operators simplifies the proof of Proposition \ref{prop:2D}  (in the appendix), as anticipated before.

The next step is to bound the left-hand side in \eqref{eq:prop2D} from below in terms of finite-size spectral gaps.

\begin{figure}[t]
\begin{center}
\includegraphics[scale=.6]{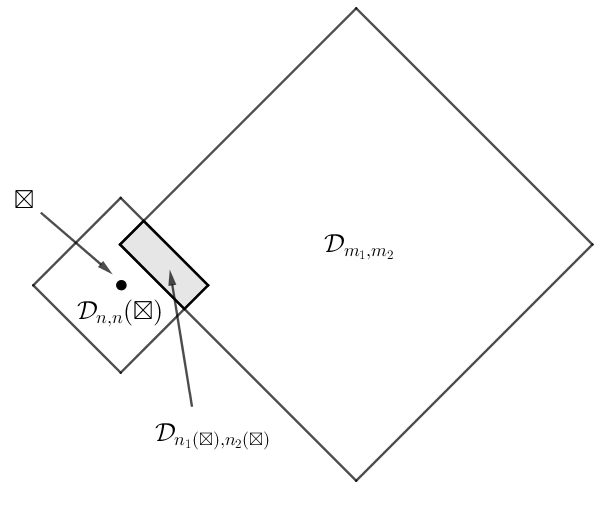}
\end{center}
\caption{We schematically depict a key geometric property of the rhomboidal patches: The (shaded) intersection of the large rhomboid $\curly{D}_{m_1,m_2}$ with the rhomboidal patch $\curly{D}_{n,n}(\boxtimes)$ centered at the plaquette $\boxtimes$ is again a rhomboid, call it $\curly{D}_{n_1(\boxtimes),n_2(\boxtimes)}$. Its sidelengths satisfy $0\leq n_i(\boxtimes)\leq n$ ($i=1,2$). Notice that if $\boxtimes$ lies well within $\curly{D}_{m_1,m_2}$, then $n_1(\boxtimes)=n_2(\boxtimes)=n$
These observations are important because the deformed patch operator $B_{n,\boxtimes}=\sum_{\Box\in P_{n,\boxtimes}} c(d_\boxtimes(\Box)) h_\Box$ actually acts only on plaquettes $\Box\in \curly{D}_{n_1(\boxtimes),n_2(\boxtimes)}$. (Recall that $P_{n,\boxtimes}:=\curly{D}_{n,n}(\boxtimes)\cap \curly{D}_{m_1,m_2}=\curly{D}_{n_1(\boxtimes),n_2(\boxtimes)}$.) .}
\label{fig:intersect}
\end{figure}

The statement takes into account how much $B_{n,\boxtimes}$ actually overlaps with the region of interest $\curly{D}_{m_1,m_2}$ (only the overlap contributes due to \eqref{eq:intersect}). Here we also use a \emph{key geometric property of rhomboids}: The intersection $\curly{D}_{n_1,n_2}(\boxtimes)\cap \curly{D}_{m_1,m_2}$ is again a rhomboid $\curly{D}_{n_1(\boxtimes),n_2(\boxtimes)}$, with smaller sidelengths $0\leq n_i(\boxtimes)\leq n$ ($i=1,2$). This is schematically shown in Figure \ref{fig:intersect} (We have $n_1(\boxtimes)=n_2(\boxtimes)=n$ well inside the bulk of the system; $n_i(\boxtimes)<n$ occurs near the boundary of $\curly{D}_{m_1,m_2}$.)

\be{lm}[Local gaps]\label{lm:n1n2}
Let $n$ be an even integer. We have the operator inequality
$$
B_{n,\boxtimes}^2 \geq c(n/2) \gam^{2D}_{n_1(\boxtimes),n_2(\boxtimes)} B_{n,\boxtimes}
$$
for appropriate integers $0\leq n_i(\boxtimes)\leq n$ ($i=1,2$) that are determined by the requirement that $P_{n,\boxtimes}\subset \curly{D}_{m_1,m_2}$ is a translated copy of $\curly{D}_{n_1(\boxtimes),n_2(\boxtimes)}$.
\e{lm}

\be{proof}
Define the (undeformed) patch operator
$$
A_{n,\boxtimes} :=\sum_{\Box\in P_{n,\boxtimes}}  h_\Box
$$
and write $\gam_{n,\boxtimes}$ for its spectral gap. By the same argument as for Proposition \ref{prop:gap}, i.e., by frustration-freeness and $\min_{1\leq r\leq n/2} c(r)=c(n/2)$, we get
$$
B_{n,\boxtimes}^2 \geq c(n/2) \gam_{n,\boxtimes} B_{n,\boxtimes}.
$$

It remains to show that
\beq\label{eq:gapeq}
\gam_{n,\boxtimes}=\gam^{2D}_{n_1(\boxtimes),n_2(\boxtimes)}
\eeq
holds for the appropriate $0\leq n_i(\boxtimes)\leq n$ ($i=1,2$). 

We consider the patches $P_{n,\boxtimes}$ more closely; recall their Definition \ref{defn:patchdefn} (i). 

The key geometric fact is that $P_{n,\boxtimes}$ is always a rhomboid. This is true even if the shifted copy of $\curly{D}_{n,n}$ that it comes from is not fully contained in $\curly{D}_{m_1,m_2}$; see Figure \ref{fig:intersect}. Indeed, for every $P_{n,\boxtimes}$, there exist integers $n_1(\boxtimes),n_2(\boxtimes)$ with $0\leq n_i\leq n$ ($i=1,2$) such that
$$
P_{n,\boxtimes}=\curly{D}_{n_1(\boxtimes),n_2(\boxtimes)}(\boxtimes)
$$
where the right-hand side is (by definition) an appropriately translated copy of $\curly{D}_{n_1(\boxtimes),n_2(\boxtimes)}$. 
 
This implies that
$$
A_{n,\boxtimes} =\sum_{\Box\in P_{n,\boxtimes}}  h_\Box=\sum_{\Box\in \curly{D}_{n_1(\boxtimes),n_2(\boxtimes)}(\boxtimes)}  h_\Box
$$
is unitarily equivalent (via translation) to the operator $H^{2D}_{n_1,n_2}$ defined in \eqref{eq:H2Ddefn}. This implies \eqref{eq:gapeq} and hence Lemma \ref{lm:n1n2}.
\e{proof}

\subsection{Step 3: Reduction to ``macroscopic'' gaps}
We apply Lemma \ref{lm:n1n2} to the result of Proposition \ref{prop:2D} and get
\beq\label{eq:up}
\begin{aligned}
\l(H^{2D}_{m_1,m_2}\r)^2+\beta_n H^{2D}_{m_1,m_2} 
\geq \al_n c(n/2) \sum_{\boxtimes \in\curly{P}} \gam^{2D}_{n_1(\boxtimes),n_2(\boxtimes)}B_{n,\boxtimes},
\end{aligned}
\eeq
with $0\leq n_1(\boxtimes),n_2(\boxtimes)\leq n$. 

In step 3, we bound the right-hand side in \eqref{eq:up} in terms of macroscopic spectral gaps via the following lemma. Recall that $\curly{D}_{n,n}(\Box)$ denotes the rhomboid $\curly{D}_{n,n}$ centered at the plaquette $\Box$. For any fixed plaquette $\Box$, let 
\beq\label{eq:sigmandefn}
\sigma_n:=\sum_{\substack{\Box' \in \curly{D}_{n,n}(\Box)}}  c(d_{\Box}(\Box')).
\eeq
An elementary computation shows that
\beq\label{eq:sigmaeq}
\sigma_n=5c(1)+\sum_{r=2}^{n/2} c(r) (16r-12)
\eeq
independently of $\Box$. 

\be{lm}[Reduction to macroscopic gaps]\label{lm:macro}
We have the operator inequality
\beq
\begin{aligned}
\sum_{\boxtimes \in\curly{P}} \gam^{2D}_{n_1(\boxtimes),n_2(\boxtimes)}B_{n,\boxtimes}
\geq \frac{\sigma_n}{4}\l(\min_{n/2\leq l_1,l_2\leq n}\gam^{2D}_{l_1,l_2}\r)H^{2D}_{m_1,m_2}.
\end{aligned}
\eeq
\e{lm}

\be{proof}
From \eqref{eq:2DBdefn}, we get
\beq\label{eq:first}
\sum_{\boxtimes \in\curly{P}} \gam^{2D}_{n_1(\boxtimes),n_2(\boxtimes)}B_{n,\boxtimes}
= \sum_{\Box\in\curly{D}_{m_1,m_2}}h_\Box\sum_{\substack{\boxtimes \in\curly{P}:\\ \Box\in P_{n,\boxtimes}}} \gam^{2D}_{n_1(\boxtimes),n_2(\boxtimes)} c(d_{\boxtimes}(\Box)).
\eeq
Fix $\Box\in\curly{D}_{m_1,m_2}$. We consider the second sum more closely. Since $\gam^{2D}_{n_1(\boxtimes),n_2(\boxtimes)}>0$ and $c(d_{\Box}(\boxtimes))>0$, we have
\beq\label{eq:second}
\begin{aligned}
\sum_{\substack{\boxtimes \in\curly{P}:\\ \Box\in P_{n,\boxtimes}}} \gam^{2D}_{n_1(\boxtimes),n_2(\boxtimes)} c(d_{\boxtimes}(\Box))
\geq& \sum_{\substack{\boxtimes \in\curly{D}_{m_1,m_2}:\\ \Box\in P_{n,\boxtimes}}} \gam^{2D}_{n_1(\boxtimes),n_2(\boxtimes)} c(d_{\boxtimes}(\Box))\\
=&\sum_{\substack{\boxtimes  \in P_{n,\Box}}} \gam^{2D}_{n_1(\boxtimes),n_2(\boxtimes)} c(d_{\Box}(\boxtimes)).
\end{aligned}
\eeq
Here we used that $d_{\boxtimes}(\Box)=d_{\Box}(\boxtimes)$ if $\Box,\boxtimes\in \curly{D}_{m_1,m_2}$. 

Since $c(d_{\Box}(\boxtimes))>0$ depends monotonically on $d_{\Box}(\boxtimes)$, the sums in \eqref{eq:second} are minimal when $\Box\in\curly{D}_{m_1,m_2}$ is equal to one of the corners of $\curly{D}_{m_1,m_2}$, e.g., 
$$
\boxdot:=Q(f_2/2).
$$
In that case, $n_1(\boxdot)=n_2(\boxdot)=n/2$ and so $P_{n,\boxdot}=\curly{D}_{n/2,n/2}$. For the next step, observe that $n/2\leq n_1(\boxtimes),n_2(\boxtimes)\leq n$ holds for every $\boxtimes  \in \curly{D}_{n/2,n/2}$ by elementary geometry.
Hence,
$$
\begin{aligned}
\sum_{\substack{\boxtimes  \in P_{n,\Box}}} \gam^{2D}_{n_1(\boxtimes),n_2(\boxtimes)} c(d_{\Box}(\boxtimes))
\geq&\sum_{\substack{\boxtimes  \in P_{n,\boxdot}}} \gam^{2D}_{n_1(\boxtimes),n_2(\boxtimes)} c(d_{\boxdot}(\boxtimes))\\
\geq& \l(\min_{n/2\leq l_1,l_2\leq n}\gam^{2D}_{l_1,l_2}\r)\sum_{\substack{\boxtimes  \in \curly{D}_{n/2,n/2}}}  c(d_{\boxdot}(\boxtimes))\\
\geq& \frac{1}{4}\l(\min_{n/2\leq l_1,l_2\leq n}\gam^{2D}_{l_1,l_2}\r)\sum_{\substack{\boxtimes  \in \curly{D}_{n,n}(\boxdot)}}  c(d_{\boxdot}(\boxtimes))\\
=& \frac{\sigma_n}{4}\l(\min_{n/2\leq l_1,l_2\leq n}\gam^{2D}_{l_1,l_2}\r).
\end{aligned}
$$
In the second-to-last step, we used that $\curly{D}_{n/2,n/2}$ forms exactly one quadrant of $ \curly{D}_{n,n}(\boxdot)$. From the definition of the distance $d_\boxdot$, this implies that it contributes precisely $1/4$ of the sum over $\boxtimes  \in \curly{D}_{n,n}(\boxdot)\setminus\{\boxdot\}$, and for the center $\boxdot$ we use the estimate $c(d_{\boxdot}(\boxdot))=c(1)\geq c(1)/4$.

Since $h_\Box\geq 0$, we may apply this estimate to \eqref{eq:first} and recall \eqref{eq:H2Ddefn} to conclude Lemma \ref{lm:macro}.
\e{proof}

\subsection{Step 4: The choice of the coefficients}
In this section, we prove Theorem \ref{thm:2D}. 

We fix two integers $m_1,m_2$ and we fix an even integer $n$ such that $2\leq n\leq \min\{m_1/2,m_2/2\}$. We apply Lemma \ref{lm:macro} to \eqref{eq:up} and find
$$
\l(H^{2D}_{m_1,m_2}\r)^2
\geq \frac{\al_n c(n/2) \sigma_n}{4}\l(\min_{l_1,l_2\leq [n/2,n]\cap \Z}\gam^{2D}_{l_1,l_2}-\frac{4\beta_n}{\al_n c(n/2) \sigma_n}\r)H^{2D}_{m_1,m_2}.
$$
Since $H^{2D}_{m_1,m_2}$ is frustration-free by Proposition \ref{prop:2Dcg}, this yields
$$
\gam_{m_1,m_2}^{2D}\geq 
\frac{\al_n c(n/2) \sigma_n}{4}\l(\min_{l_1,l_2\leq [n/2,n]\cap \Z}\gam^{2D}_{l_1,l_2}-G(n)\r).
$$
where
$$
G(n):=\frac{4\beta_n}{c(n/2) \al_n  \sigma_n}.
$$
By Proposition \ref{prop:2Dcg}, specifically \eqref{eq:prop2Dgap}, we have
$$
C_1^{2D}\gam^{2D}_{m_1,m_2}\leq \gam_{\curly{R}_{m_1,m_2}}\leq C_2^{2D} \gam^{2D}_{m_1,m_2}
$$
and so
\beq
\gam_{\curly{R}_{m_1,m_2}}\geq 
C_1^{2D}\frac{\al_n c(n/2) \sigma_n}{4C_2^{2D}}\l(\min_{l_1,l_2\leq [n/2,n]\cap \Z}\gam_{\curly{R}_{l_1,l_2}}-C_2^{2D} G(n)\r).
\eeq
It remains to bound $G(n)$ by a judicious choice of coefficients.

\be{lm}[Choice of coefficients]\label{lm:G}
For $1\leq r\leq n/2$, define
\beq\label{eq:cdefn}
c(r):=n^{3/2}+\l(\frac{n}{2}\r)^2-r^2
\eeq
Then $
G(n)\leq  n^{-3/2}.$
\e{lm}

Note that the coefficients defined in \eqref{eq:cdefn} satisfy Assumption \ref{ass:c2D}. Hence, Lemma \ref{lm:G} implies Theorem \ref{thm:2D} with the constants 
\beq\label{eq:see}
C_1:=C_1^{2D}\frac{\al_n c(n/2) \sigma_n}{4C_2^{2D}},\qquad C_2:=C_2^{2D}
\eeq
We recall that $C_1^{2D}$ and $C_2^{2D}=4\lam_{\max,eff}$ were defined in the proof of Proposition \ref{prop:2Dcg}. It remains to give the

\be{proof}[Proof of Lemma \ref{lm:G}]
Recall Definitions \eqref{eq:albetadefn} of $\al_n,\beta_n$ and formula \eqref{eq:sigmaeq} for $\sigma_n$. After some elementary manipulations, we use \eqref{eq:cdefn} to get
$$
\begin{aligned}
G(n)
=&4\frac{(2n-1)c(n/2)^2+\sum_{r=2}^{n/2}(4r-5)\l(c(r)-c(r-1)\r)^2}{c(n/2)\l(5c(1)+\sum_{r=2}^{n/2} c(r) (16r-12)\r)}\\
\leq&  n^{-3/2}. 
\end{aligned}
$$
This proves Lemma \ref{lm:G} and hence Theorem \ref{thm:2D}.
\e{proof}

\be{appendix}

\section{Knabe-type bound in 1D}
In this appendix, we prove Proposition \ref{prop:rewrite}. We begin by computing $B_{n,l}^2$. Using the fact that $h_j^2=h_j$, we find
\beq
\label{eq:square}
\begin{aligned}
B_{n,l}^2
=&\sum_{j=l}^{l+n-2} (c_{j-l})^2 h_j
+\sum_{j=l}^{l+n-3} c_{j-l} c_{j+1-l}\{h_j,h_{j+1}\}\\
&+\sum_{j=l}^{l+n-4} \sum_{j'=j+2}^{l+n-2}c_{j-l}c_{j'-l}\{h_j,h_{j'}\}.
\end{aligned}
\eeq
We use this identity to rewrite the left-hand side of the claim in Propositon \ref{prop:rewrite}. After interchanging the order of summation, we get
\beq\label{eq:fubini1}
\begin{aligned}
\sum_{l=1}^{m+1} B_{n,l}^2
=&\l(\sum_{j=0}^{n-2} c_j^2\r) H_m+\l(\sum_{j=0}^{n-3}c_jc_{j+1}\r)Q\\
&+\sum_{\substack{1\leq i,i'\leq m+1\\ 2\leq d(i,i')\leq n-2}} h_{i}h_{i'}\l(\sum_{j=0}^{n-2-d(i,i')}c_j c_{j+d(i,i')}\r).
\end{aligned}
\eeq

We employ the following ``1D Autocorrelation Lemma'', which is proved in the appendix to \cite{GM}. We recall that the collection of positive real numbers $\{c_0,\ldots,c_{n-2}\}$ satisfies Assumption \ref{ass:c}.

\be{lm}[1D Autocorrelation Lemma \cite{GM}]\label{lm:ac}
For every $0\leq x\leq n-3$, the quantity
$$
q(x)=\sum_{j=0}^{n-2-x} c_j c_{j+x}.
$$
satisfies
$$
q(x)\geq q(x+1).
$$
\e{lm}

By Lemma \ref{lm:sign}, we have $h_ih_{i'}\geq 0$ for all $d(i,i')\geq2$. Therefore, the sum indexed by $1\leq i,i'\leq m+1$ in \eqref{eq:fubini1} is taken over non-negative terms. We can thus apply Lemma \ref{lm:ac} to the coefficient of each $h_i h_{i'}$. We conclude
$$
\begin{aligned}
\sum_{l=1}^{m+1} B_{n,l}^2
\leq&
\l(\sum_{j=0}^{n-2} c_j^2\r) H_m
+\l(\sum_{j=0}^{n-3}c_jc_{j+1}\r)
\l(Q+\sum_{\substack{1\leq i,i'\leq m+1\\ 2\leq d(i,i')\leq n-2}} h_{i}h_{i'}\r)\\
\leq&
\l(\sum_{j=0}^{n-2} c_j^2\r) H_m
+\l(\sum_{j=0}^{n-3}c_jc_{j+1}\r)
\l(Q+F\r).
\end{aligned}
$$
In the last step, we used Lemma \ref{lm:sign} once more. This proves Proposition \ref{prop:rewrite}.
\qed

\section{Knabe-type bound in 2D}
In this appendix, we prove Proposition \ref{prop:2D}.

\subsection{Preliminaries}
To compute the squares of the deformed patch operators, we use the formalism from \cite{GM}.

\be{defn}\label{defn:W}
We define an equivalence relation $\sim$ among pairs of plaquettes $(\Box_1,\Box_2), (\Box_3,\Box_4)\in\curly{P}\times \curly{P}$. We call these pairs equivalent iff they differ by a lattice translation, i.e., 
$$
(\Box_1,\Box_2)\sim (\Box_3,\Box_4)\; :\Longleftrightarrow\; \exists v\in \Z^2\;\textnormal{ s.t. }\; (\Box_1+v,\Box_2+v)= (\Box_3,\Box_4).
$$
Let $n\geq 2$ be an even integer. Fix $\boxdot\in \curly{P}$ (think of this as fixing an origin). On $\curly{D}_{n,n}(\boxdot)$, we define the distance function
$$
d_n(\Box):=\min\setof{1\leq k\leq n/2}{\Box\in \curly{D}_{2k,2k}(\boxdot)}.
$$
Then we define a weight function $W_n:\curly{D}_{m_1,m_2}\times \curly{D}_{m_1,m_2}\to \R_+$ via
\beq\label{eq:Wdefn}
W_n(\Box_1,\Box_2):=\sum_{\substack{\Box_3,\Box_4\in \curly{D}_{n,n}(\boxdot)\\ (\Box_3,\Box_4)\sim (\Box_1,\Box_2)}} c(d_n(\Box_3)) c(d_n(\Box_4)).
\eeq
\e{defn}

Recall that $B_{n,\boxtimes} :=\sum_{\Box\in P_{n,\boxtimes}} c(d_\boxtimes(\Box)) h_\Box$. With these definitions and $h_\Box^2=h_\Box$, we have
\beq\label{eq:Bsquare}
\sum_{\boxtimes\in \curly{P}} B_{n,\boxtimes}^2=\sum_{\Box\in \curly{D}_{m_1,m_2}} W_n(\Box,\Box) h_\Box
+\sum_{\substack{\Box,\Box'\in \curly{D}_{m_1,m_2}\\ \Box\neq\Box'}} W_n(\Box,\Box') h_\Box h_{\Box'}.
\eeq
To see this, it is convenient to temporarily forget the restriction that $P_{n,\boxtimes}$ is intersected with $\curly{D}_{m_1,m_2}$ (e.g., one may introduce $h_\Box:=0$ for $\Box\notin \curly{D}_{m_1,m_2}$). The computation in the whole space is straightforward. Afterwards, one remembers that only terms in $\curly{D}_{m_1,m_2}$ are kept. 

As in 1D, the relevant terms in \eqref{eq:Bsquare} are the self-interactions and nearest-neighbor terms. We denote these by
$$
\begin{aligned}
W_n^{\mathrm{self}}:=&W_n(\Box,\Box),\quad\textnormal{for any }\Box \in \curly{D}_{m_1,m_2},\\
W_n^{\mathrm{edge}}:=&W_n(\Box,\Box')\quad \textnormal{for any } \Box,\Box'\in \curly{D}_{m_1,m_2} \textnormal{ sharing an edge},\\
W_n^{\mathrm{corner}}:=&W_n(\Box,\Box')\quad \textnormal{for any } \Box,\Box'\in \curly{D}_{m_1,m_2} \textnormal{ sharing a corner}.
\end{aligned}
$$
Notice that the right-hand sides do not depend on the particular plaquettes that they are evaluated on, since all choices related by lattice translations or rotations. In the following lemma, we compute these three quantities.

\be{lm}\label{lm:W}
Let $n\geq 2$ be an even integer. We have
$$
\begin{aligned}
W_n^{\mathrm{self}}=&5c(1)^2+\sum_{r=2}^{n/2} c(r)^2 (16r-12),\\
W_n^{\mathrm{edge}}=&W_n^{\mathrm{corner}}=2c(1)^2+\sum_{r=2}^{n/2} \l(c(r)^2 (8r-6) +c(r)c(r-1)(8r-10)\r).
\end{aligned}
$$
\e{lm}

\be{rmk}\label{rmk:W}
The fact that $W_n^{\mathrm{edge}}=W_n^{\mathrm{corner}}$ is critical; it is the reason why we chose the rhomboidal patches. (Compare also eq.\ (54) in \cite{GM}.) The fact that $d_n$ changes its value only at every second step is also crucial. 
\e{rmk}

\be{proof}
By \eqref{eq:Wdefn}, we have
$$
W_n^{\mathrm{self}}=W_n(\Box,\Box)=\sum_{\Box\in \curly{D}_{n,n}} c(d_n(\Box))^2=5c(1)^2+\sum_{r=2}^{n/2} c(r)^2 (16r-12).
$$
For $W_n^{\mathrm{edge}}$ and $W_n^{\mathrm{corner}}$, we use induction. The induction base is
$$
W_2^{\mathrm{edge}}=W_2^{\mathrm{corner}}=2c(1)^2.
$$
The claim then follows from the recursion relations
$$
W_n^\#=W_{n-2}^\# +c\l(\frac{n}{2}\r)c\l(\frac{n}{2}-1\r) (4n-10)+\l(c\l(\frac{n}{2}\r)\r)^2 (4n-6),
$$
where $\#\in \{\mathrm{edge},\mathrm{corner}\}$. This proves Lemma \ref{lm:W}.
\e{proof}

\subsection{Proof of Proposition \ref{prop:2D}}
We write $\Box\leftrightarrow \Box'$ to express that $\Box,\Box'$ are distinct and share an edge or a corner, and we write $\Box\not\leftrightarrow \Box'$ to express that $\Box,\Box'$ are distinct and do not share an edge or a corner.

On the one hand, we can apply Lemma \ref{lm:W} to \eqref{eq:Bsquare}. This gives
\beq\label{eq:Bsquare'}
\begin{aligned}
\sum_{\boxtimes\in \curly{P}} B_{n,\boxtimes}^2
=&W_n^{\mathrm{self}} H^{2D}_{m_1,m_2}
+W_n^{\mathrm{edge}} \sum_{\substack{\Box,\Box'\in \curly{D}_{m_1,m_2}\\ \Box\leftrightarrow \Box'}} h_\Box h_{\Box'}+\sum_{\substack{\Box,\Box'\in \curly{D}_{m_1,m_2}\\ \Box\not\leftrightarrow \Box'}} W_n(\Box,\Box') h_\Box h_{\Box'}.
\end{aligned}
\eeq
On the other hand, we can square the Hamiltonian $H^{2D}_{m_1,m_2}$ to get
\beq\label{eq:Hsquare}
\l(H^{2D}_{m_1,m_2}\r)^2=H^{2D}_{m_1,m_2}+\sum_{\substack{\Box,\Box'\in \curly{D}_{m_1,m_2}\\ \Box\neq\Box'}} h_\Box h_{\Box'}.
\eeq
Recall definitions \eqref{eq:albetadefn} -- in particular $\al_n=1/W_n^{\mathrm{edge}}$. We subtract $\al_n$ times \eqref{eq:Bsquare'} from \eqref{eq:Hsquare} (to cancel the terms with $\Box\leftrightarrow \Box'$) and obtain
$$
\l(H^{2D}_{m_1,m_2}\r)^2\geq  \al_n \sum_{\boxtimes \in\curly{P}} B_{n,\boxtimes}^2 -\beta_n H^{2D}_{m_1,m_2}
+\sum_{\substack{\Box,\Box'\in \curly{D}_{m_1,m_2}\\ \Box\not\leftrightarrow\Box'}} \l(1-\al_n W_n(\Box,\Box')\r) h_\Box h_{\Box'}.
$$
The last term is non-negative by the 2D Autocorrelation Lemma \ref{lm:2Dac} below and Proposition \ref{prop:2D} is proved.
\qed

\subsection{2D Autocorrelation Lemma}
\be{lm}\label{lm:2Dac}
Let $n\geq 2$ be an even integer and let $\{c(1),\ldots,c(n/2)\}$ satisfy Assumption \ref{ass:c2D}. Let $\Box_1,\Box_2\in \curly{D}_{n,n}$ be distinct. Then
$$
W_n(\Box_1,\Box_2)\leq W_n^{\mathrm{edge}}.
$$
\e{lm}

\be{proof}
Fix an even integer $n\geq 2$ and two distinct plaquettes $\Box_1,\Box_2\in \curly{D}_{n,n}$. We follow the strategy in \cite{GM}; our situation is simpler because orientation does not matter.

 We express the function  $\Box\mapsto c(d_n(\Box))$ in terms of its level sets. Since the coefficients $\{c(1),\ldots,c(n/2)\}$ are positive and non-increasing, there exist nonnegative numbers $\{t_1,\ldots,t_{n/2}\}$ such that
$$
c(d_n(\Box))=\sum_{j=1}^{n/2} t_j b_j(\Box),\qquad b_j(\Box):=\ind_{\setof{\Box}{d_n(\Box)\leq j}},
$$
where $\ind_X$ is an indicator function for the set $X$. 

Recall that $d_n$ and $W_n$ are defined in terms of a rhomboid $\curly{D}_{n,n}(\boxdot)$ centered at some fixed $\boxdot\in \curly{P}$. We abbreviate this rhomboid by $\curly{D}_n^0:=\curly{D}_{n,n}(\boxdot)$. By interchanging the order of summation, we get
\beq\label{eq:via}
\begin{aligned}
W_n(\Box_1,\Box_2)
= \sum_{\substack{\Box_3,\Box_4\in \curly{D}_n^0\\ (\Box_3,\Box_4)\sim (\Box_1,\Box_2)}} c(d_n(\Box_3)) c(d_n(\Box_4))
= \sum_{j,j'=1}^{n/2} t_j t_{j'} R_{j,j'}(\Box_1,\Box_2)
\end{aligned}
\eeq
with the purely geometric quantity
$$
R_{j,j'}(\Box_1,\Box_2):=\sum_{\substack{\Box_3,\Box_4\in \curly{D}_n^0\\ (\Box_3,\Box_4)\sim (\Box_1,\Box_2)}} b_j(\Box_3)b_{j'}(\Box_4).
$$
From now on, we identify $\curly{D}_n^0$ with the plaquettes of a rhomboidal subset of $\Z^2\subset \R^2$, i.e., we take the plaquettes to be squares of sidelength $1$.

Given the two distinct plaquettes $\Box_1,\Box_2\in \curly{D}_n^0$, there is a plaquette in $\curly{D}_n^0$ which (a) shares an edge with $\Box_1$ and (b) is closer to $\Box_2$ than $\Box_1$ is (``closer'' meaning that their centers are closer as points in $\R^2$). We write $\boxtimes_2$ for this plaquette. (There may be up to two such plaquettes. If there are two, we take $\boxtimes$ to be the one of the two that is closer to the center of $\curly{D}_n^0$. If they are equally close to the center, we take either one.)

We will show that for any $1\leq j,j'\leq n/2$,
\beq\label{eq:boxclaim}
R_{j,j'}(\Box_1,\Box_2)\leq R_{j,j'}(\Box_1,\boxtimes_2).
\eeq
Since $t_j\geq 0$, we can apply this bound to \eqref{eq:via} to conclude the claim.

We now prove \eqref{eq:boxclaim}. Let $T$ be the translation of $\R^2$ that maps $\Box_1$ to $\Box_2$ (i.e., $T(v)=v+\Box_2-\Box_1$). Then $(\Box_1,\Box_2)\sim (\Box_3,\Box_4)$ implies $\Box_4=T(\Box_3)$ and so
$$
\begin{aligned}
R_{j,j'}(\Box_1,\Box_2)
=&\l|\setof{(\Box_3,\Box_4)}{\Box_3\in\curly{D}^0_{2j},\quad \Box_4\in\curly{D}^0_{2j'},\quad \Box_4=T(\Box_3)}\r|\\
=&\l|\setof{(T^{-1}(\Box_4),\Box_4)}{T^{-1}(\Box_4)\in\curly{D}^0_{2j},\quad \Box_4\in\curly{D}^0_{2j'}}\r|\\
=&\l|T(\curly{D}^0_{2j})\cap \curly{D}^0_{2j'}\r|,
\end{aligned}
$$ 
where $|\cdot|$ denotes cardinality. The same argument gives
$$
R_{j,j'}(\Box_1,\boxtimes_2)=\l|\tilde T(\curly{D}^0_{2j})\cap \curly{D}^0_{2j'}\r|
$$
for $\tilde T$ the translation  that maps $\Box_1$ to $\boxtimes_2$ (i.e., a shift by $1$ lattice unit in $\Z^2$). Without loss of generality, suppose that $j\leq j'$, so $\curly{D}^0_{2j}\subset \curly{D}^0_{2j'}$. 

If $j<j'$, then $\tilde T(\curly{D}^0_{2j})\subset \curly{D}^0_{2j'}$ still and so 
$$
R_{j,j'}(\Box_1,\boxtimes_2)=|\curly{D}^0_{2j'}|\geq \l|T(\curly{D}^0_{2j})\cap \curly{D}^0_{2j'}\r|=R_{j,j'}(\Box_1,\Box_2),
$$
proving \eqref{eq:boxclaim} in that case.

If $j=j'$, then $\l|T(\curly{D}^0_{2j})\cap \curly{D}^0_{2j'}\r|\leq |\curly{D}^0_{2j'}|-(4j'-1)$ with equality iff $T$ is a shift by $\{(\pm 1,0),(0,\pm 1),(\pm 1/\sqrt{2},\pm 1/\sqrt{2})\}$ (recall that we consider the lattice as a subset of $\Z^2$). In particular, equality holds for $\tilde T$; hence
$$
\begin{aligned}
R_{j,j'}(\Box_1,\Box_2)=&\l|T(\curly{D}^0_{2j})\cap \curly{D}^0_{2j'}\r|\\
\leq& |\curly{D}^0_{2j'}|-(4j'-1)=\l|\tilde T(\curly{D}^0_{2j})\cap \curly{D}^0_{2j'}\r|
=R_{j,j'}(\Box_1,\boxtimes_2).
\end{aligned}
$$
We have therefore shown \eqref{eq:boxclaim} for all $j,j'$. This finishes the proof of Lemma \ref{lm:2Dac}.
\e{proof}

\end{appendix}

{\footnotesize 
\paragraph{Acknowledgments.}
We thank Matthew Cha, Elliott H.\ Lieb, Ramis Movassagh, Bruno Nachtergaele, Jeff Schenker, Tom Spencer and especially David Gosset for helpful comments. ML is grateful to the Institute for Advanced Study for its hospitality during the 2017-2018 academic year. This material is based upon work supported by the National Science Foundation under Grant No.\ DMS - 1638352. 
\par}

\begin{footnotesize}

\end{footnotesize}
\end{document}